\documentclass[useAMS,usenatbib]{mn2e}
\usepackage{graphicx}
\usepackage{amssymb}
\usepackage{lscape}

\usepackage{longtable}                                    %
\newcounter{qub}                                          %
\newcommand{\qq}{\addtocounter{qub}{1}\arabic{qub}}       %

\newcommand{\apj}{ApJ}

\newcommand{\aap}{A\&A}

\newcommand{\mnras}{MNRAS}

\newcommand{\kms}{km\,s$^{-1}$}

\newcommand{\HI}{\ion{H}{i}}
\newcommand{\HII}{\ion{H}{ii}}

\newcommand{\sunn}{$_{\odot}$}

\DeclareRobustCommand{\ion}[2]{%
\relax\ifmmode
\ifx\testbx\f
{\mathrm{#1\,\textsc{#2}}}\else
{\mathrm{#1\,\mathsc{#2}}}\fi
\else\textup{#1\,{\mdseries\textsc{#2}}}%
\fi}

\title[Void galaxies in the nearby Universe. I. Sample] 
{Void galaxies in the nearby Universe. I. Sample  description}
\author[S.A. Pustilnik, A.L. Tepliakova, D.I. Makarov]
{S.A. Pustilnik,$^{1}$\thanks{E-mail: sap@sao.ru (SAP),dim@sao.ru (DIM)}
A.L. Tepliakova,$^{1}$ D.I. Makarov$^{1}$ \\
\rule{-4pt}{20pt}
$^1$ Special Astrophysical Observatory of RAS, Nizhnij Arkhyz,
 Karachai-Circassia 369167, Russia}

\begin{document}

\label{firstpage}

\date{Accepted 2018 August ??. Received 2018 July 30}

\pagerange{\pageref{firstpage}--\pageref{lastpage}} \pubyear{2018}

\maketitle

\begin{abstract}
The main goal of this work is to form a large, deep and representative
sample of dwarf galaxies residing in voids  of the nearby Universe.
The formed sample is the basement for the comprehensive mass study of
the galaxy content, their evolutionary status, clustering and dynamics
with respect to their counterparts residing in more typical, denser
regions and for study of void small-scale substructures. We present
25 voids over the entire sky within 25 Mpc from the Local Group.
They are defined as groups of lumped empty spheres bounded by `luminous'
galaxies with the absolute K-band magnitudes brighter than --22.0.
The identified void regions include the Local Void and other known
nearby voids. The nearest nine voids occupy a substantial part
of the Local Volume. Of the total number of 6792 cataloged galaxies
in the considered volume, 1354 objects fall into 25 nearby voids.
Of this general void galaxy sample, we separate the sub-sample of
`inner' void galaxies, residing deeper in voids, with distances to
the nearest luminous galaxy $D_{\rm NN} \geq 2.0$ Mpc. The `inner'
galaxy sample includes 1088 objects, mostly dwarfs with $M_{\rm B}$
distribution peaked near --15.0 and extending down to --7.5 mag.
Of them, 195 fall in the Local Volume (space within R=11 Mpc). We present
the general statistical properties of this Nearby Void Galaxy sample
and discuss the issues related to the sample content and the prospects of its use.
\end{abstract}


\section{Introduction}
\label{sec:intro}

Voids as elements of the large scale structure fill $\sim$70 per cent
of the modern Universe volume and represent its the lowest density regions.
While the void galaxies comprise only a small fraction ($\sim$15 per cent) of
the whole galaxy population, their properties can give the important
indications on the processes at the earlier epochs in the Universe.
The study of voids and properties of galaxies residing in them remains one of
the actual and important field of extragalactic and cosmological research
 \citep{Weygaert16,AragonCalvoSzalay2013}.

Massive observational studies of voids are mainly based 
on the modern surveys such as SDSS and ALFALFA
\citep{GG99,Rojas04,Rojas05,Hoyle05,Hoyle15,Patiri06,Kreckel11,Kreckel12,Pan2012}
as well on the earlier surveys (Las Campanas, 2dFGRS, 6dF, CfA).
As a matter of fact, the great majority of the previous works dealt with
distant large voids ($D \sim$ 80--200 Mpc). The common limit of used
spectroscopic redshift surveys ($r \lesssim 17.8$, or $B\lesssim 18.5$) and
the applied requirements of statistical completeness  resulted in the strong bias
of the studied void galaxies to the upper part of the luminosity function
(LF), e.i., $M_{\rm B} \sim -16 $ to $-20$.
The comparison of these most luminous void galaxies reveals only relatively
weak differences of properties with similar galaxies in the adjacent walls
(or denser environments).

In the recent series of papers we suggested a complementary approach aimed
to study galaxy properties in a nearby Lynx-Cancer void, with the emphasis
on the least massive void representatives
\citep{PaperI,PaperII,void_LSBD,PaperIV,Paper6,Paper7,DDO68}.
These studies resulted in the firm conclusions on a systematically lower
metallicity of gas and a higher \HI-mass fraction of the void galaxy
population in comparison to the similar galaxies in denser environments.
These results imply that void galaxies in average evolve significantly
slower than their counterparts in typical groups. Besides, it appeared
that the substantial fraction of the faintest LSB dwarfs in the Lynx-Cancer
void (mostly those with $M_{\rm B} \gtrsim -13.5$ mag) show the very low O/H,
extremely high gas mass fraction and blue colour of stars at the galaxy
periphery.
These properties indicate the atypical young evolutionary status of
such LSB dwarfs \citep{J0926,CP13,J0015,Paper7,UGC3672A}.

To study the fainter part of void galaxies, one needs first of all,
to use the samples of sufficiently close objects. Since the available
wide-field spectral
surveys of galaxies have the limiting apparent magnitudes no fainter than
(or close to) $B_{\rm tot} = $ 18.5 -- 19 mag,
in order to deal with dwarfs as faint as $M_{\rm B} = -13$ mag,
we need in a galaxy sample with the distance moduli less than $\sim 32$ mag,
which implies the distances of $D < 25$~Mpc.

Unlike the most of the last decade studies of voids and their galaxies, 
we limit our task by the separation of large `simply connected' `empty'
volumes devoid of galaxies brighter than $M_{\rm K} = M_{\rm K}^{*}+2$. Here
$M_{\rm K}^{*}$ = -24.0 is the
characteristic absolute magnitude of the $K$-band luminosity function.

It is worth to mention that several nearest voids were identified by
\citet{Fairall98} and shown in his Atlas. However, over the past 20 years,
the number of known nearby galaxies increased substantially.
Therefore it is necessary to fulfill a new analysis of distribution and
galaxy population of nearby voids. Recently, \citet{Elyiv2013} studied
the properties of voids in the local Universe ($R < 50$~Mpc).
They found many empty spheres interconnected into elongated structures
(tunnels). While this work allowed the authors to probe the most rarefied
space regions, their original selection causes significant limitations for
void parameters and void galaxy sample.
Namely, their voids were constructed with the sample of galaxies delineating
voids to be of  low luminosity: $M_{\rm K} < -18.4$ mag.
This limit is 5.6 magnitudes fainter than the characteristic absolute
magnitude of galaxy Luminosity Function $M_{\rm K}^{*} = -24.0$ mag
\citep{Hill2010}.

Galaxies with $L$  somewhat lower than the luminosity $L^{*}$  are commonly
used to define void boundaries. In numerical cosmological simulations,
the masses of Dark Matter (DM) halos, delineating voids, usually correspond to
objects with the luminosity $\sim$2--3 times smaller than the characteristic
one. For example, in simulations by \citet{AragonCalvoSzalay2013}, voids are
determined by Cold DM halos corresponding to $M_{\rm r} \lesssim -20.7$ mag,
what is $\sim 0.8$ mag (or a factor of 2) fainter than $M_{\rm r}^{*} = -21.5$ mag
\citep{Montero09,Hill2010}. The galaxies with the intermediate luminosities,
of $M_{\rm r} \lesssim -18.7$ define smaller subvoids
in void substructures. Respectively, CDM halos of galaxies with
$M_{\rm r} \lesssim -16.7$ define even smaller subsubvoids in those subvoids.
In this context, the voids of \citet{Elyiv2013}
delineated by galaxies of 170 times less luminous than $L_{\rm K}^{*}$,
appear systematically smaller with respect of the previous studies.
Majority of them should represent the fine structure of the rarefied regions
or 'small' cavities in much larger regions with the density above the mean
one.

In the recent paper by \citet{Hidding16} several voids in the nearby
Universe were identified via a sophisticated analysis of the density
field and galaxy peculiar velocities. However, they are shown only as
illustrations in the significantly larger cell  than considered here
(namely, the cube with the side of $\sim$120~Mpc).

The goal of this work is to select large voids in the vicinity of the
Local Volume (namely, within $R = 25$~Mpc). The relative proximity of
galaxy population in these nearby voids allows us with the use of
the deep wide-field redshift surveys (like SDSS, 2dFGRS, ALFALFA)
to find and study many intrinsically faint dwarfs with $M_{\rm B}$
down to --12 mag and even to --10 mag. Despite the lack of the
statistical completeness, the understanding of properties of the
least massive void dwarfs can provide us with the important clues
on the origin and evolution of the majority of galaxy population
within the major part of the Universe volume occupied by voids.

Apart the void galaxy evolution issues
\citep[e.g.][]{Hoeft06,Hahn06,Hahn07,Hahn09},
the nearby void sample should
be useful for the study of voids themselves. According to
\citet{AragonCalvoSzalay2013}, the internal filaments of voids
with scales of\, $\lesssim$~2~Mpc are still in the linear regime
of mass assembly and therefore reflect the initial mass distribution.
This, in less extent can relate to the low mass DM halos and their dwarfs.
We can probe the small-scale structure of voids and study its properties
only with a denser filling of a void volume with the test particles.
Due to the known rise of the galaxy LF for  lower luminosities
(masses), to reach a larger galaxy number density, one needs to
go deeper for  void galaxy selection  in their luminosity or mass.

Another interesting implication of the study of the least massive objects
and their structures in voids is the expected sensitivity of filaments,
DM halos and their galaxies to the admixture of Warm Dark Matter (WDM),
since the presence of the WDM component can wash out the small-scale
disturbances and thus limit the existence of the smallest galaxies
and their structures \citep[e.g.,][]{Angulo2013}.

The rest of the paper is designed as follows. In Section \ref{sec:voids}
we describe in detail all main procedures to construct `empty' regions as
defined by the `luminous' galaxy sample.
In Section \ref{sec:voidgal} we describe the selection of void galaxies.
Section \ref{sec:results} presents
the list with the main geometrical parameters of the separated Nearby Voids
and parameters of void galaxies taken from HyperLEDA and/or NED.
In Section \ref{sec:dis} we briefly discuss statistical properties of the
Nearby Voids' galaxies, compare them with the previous results and draw the
prospects of the galaxy sample study. We finally summarize the results of
this work and conclude in Section \ref{sec:sum}. For all distance dependent
parameters, we adopt the Hubble constant of $H_{\mathrm 0}=73$~\kms~Mpc$^{-1}$.
Due to their large volume for the printed version, the major part of
tables and illustrative materials like the finding charts of the Nearby
Voids' galaxies and the 3D presentations of the individual nearby voids,
are grouped in Appendix and presented in On-line materials.

\section{Void sample}
\label{sec:voids}

\subsection{Overview}
\label{ssec:over}

There are many void-finding methods described in the literature, ranging from
those looking for maximal empty spheres in point distibutions, either
of DM halos in simulations, or galaxies in volume-limited samples [e.g.,
\citet{Gottlober03} and references therein] to various 'advanced' schemes
looking for density minima in the DM or smoothed galaxy density distributions
[e.g., \citet{Colberg05,Hahn06,Hidding16}].
The majority of the modern methods were compared in \citet{Colberg2008}.
This comparison shows that despite the similarity of the most empty regions
found by different methods, the basic parameters of voids can differ
significantly from one method to another. One of implications of the above
conclusion is the difficulty in the attempts
to confront voids and their population separated on the distribution of
real galaxies and those found on simulations with known distribution of DM.

In our approach we approximate nearby voids by the groups of lumped empty
spheres bounded by the spatial distribution of luminous (massive) galaxies
and galaxy groups to which they belong.
This void selection is similar to the work of \citet{Patiri06}, but our
method allows the found empty spheres to lump and thus, to form
non-spherical voids, consisting of several close empty spheres.

The selection of the empty spheres, in turn, is based on Voronoi tessellation
of cells around luminous galaxies
\citep[see, e.g.,][]{MatsudaShima1984,IckeVanDeWeygaert1987,
IckeVanDeWeygaert1991,VanDeWeygaertIcke1989}.
Voronoi tessellation divides the 3D space into cells in such a way that
for each luminous galaxy there is corresponding region consisting of
all points closer to that galaxy than to any other.
Thus, the wall between the two cells consists of the points equally distant
from the two neighbouring galaxies. The nodes of the grid delineating the
cells are the local maxima of distances to the nearest galaxies, and
they are centres of the maximal empty spheres which can be inscribed
in this region of space.
The algorithm of the void selection was implemented using the
\textit{MATLAB} package.

Before the use of the sample of luminous galaxies, we apply the algorithm
of group selection described by \citet{LSCPairs,LSCTriplets,LSCGroups}.
The final sample of massive objects which delineate the empty spheres, 
includes both lonely luminous galaxies and  groups of galaxies with
the total luminosity exceeding the specific limit. The smaller galaxies
identified as the members of these `luminous groups' are excluded from the
further analysis as by definition representing non-void galaxies.

To construct voids in the considered volume, we combine several empty spheres
which obey the condition of a sufficiently small distance between the centres
of two adjacent spheres (see details below).

\subsection{Procedures of void construction}
\label{ssec:procedure}

We created the list of galaxies with the radial velocities
$V_{\rm LG} < 3500$~\kms\  relative to the Local Group (LG)
rest frame \citep{LGApex1996} using the
HyperLEDA\footnote{http://leda.univ-lyon1.fr/} database \citep{HyperLEDA}.

To minimize the effect of dust extinction in the Milky Way, we use $K$-band
magnitudes for luminous galaxies, for which the attenuation of light, even
in the Zone of Avoidance (ZOA), is a modest. Thus, the luminous galaxies with
$R < 25$~Mpc are not lost.
The absolute total magnitudes, $M_{\rm K,tot}$, of galaxies are based on
the photometry from the Two-micron All Sky Survey (2MASS) \citep{2MASS}
and the distances from the model of peculiar velocity field
in the vicinity of the Local Volume \citep{Tully2008}. In cases
when the distances are known from the accurate velocity-independent methods,
we use the latter distances (see below).
The $K$-band characteristic value of the luminosity function
is $M_{\rm K}^{*} = -24.0$ mag as adopted from \citet{Hill2010}. This is well
consistent with other $M_{\rm K}^{*}$ estimates cited in their Table~5.
We define 'luminous' galaxies as those with
$M_{\rm K} = -22.0$ mag or brighter. This threshold is 2.0 mag fainter
than $M_{\rm K}^{*}$.
These galaxies and groups around them determine void borders.

The adopted threshold in $M_{\rm K}$ at 2 mag fainter than $M_{\rm K}^{*}$
looks consistent with the $B$-band absolute magnitude $M_{\rm B} = -18.4$
for the most luminous galaxies in the nearby Lynx-Cancer void \citep{PaperI}.
The latter $M_{\rm B}$ is $\sim$2.1 mag fainter than $M_{\rm B}^{*} \sim -20.5$
according to \citet{Hill2010} and the earlier estimates. In fact, however,
we identify in the separated voids of $\sim$3 per cent subluminous galaxies with
$M_{\rm B} =$ --18.4 to --19.5 mag. So, the current Nearby Void
galaxy sample differs from the studied in the Lynx-Cancer void by the presence
of the small fraction of brighter subluminous objects.

Another important correction is related to the distance estimation
from the redshift measurements. As shown by \citet{Tully2008}, the
apparent radial velocities of galaxies in the Local Volume and the
adjacent regions are affected by the influence of the Local Supercluster
and the large Local Void.

We applied the zero order peculiar velocity correction from
\citet{Tully2008} related to the combination of motions of the Local Sheet
(LS) of galaxies in the directions from the Local Void
($L=11\degr, B=-72\degr$) and to the Local Supercluster centre (close to
the Virgo Cluster) ($L=103\degr, B=-2\degr$),
which in sum is $323\pm25$~\kms\ in the direction $\mathbf{n}_{\rm LS}$ 
with the equatorial coordinates $RA = 8^h 55^m$, $Dec = +08\degr$.

According to \citet{Tully2008}, the LG moves with respect of the
LS coordinate system with the velocity of $V = 60\pm24$~\kms\ in
the direction of $L=150\degr$, $B=+53\degr$.
Then, to apply this peculiar velocity correction to a commonly
used velocities $V_{\rm LG}$, we transform vector $\mathbf{n}_{\rm LS}$
to the LG coordinate system. This results in the motion with
$V_{\rm LS} = $357~\kms\ in the direction $L=70.8\degr$, $B=-58.5\degr$.
Thus, we adopt at zero approximation the velocity correction 
which accounts the motion of the LG with respect of the LS 
and of the LS as combination of the Local Void repulsion and the
Local Supercluster attraction,  $V_{\rm cor} = V_{\rm LG} + \Delta V$.
The correction to the galaxy radial velocity
$\Delta V = 357~\cos(\theta)$~\kms.
Here $\theta$ is the angle between the vector $\mathbf{n}_{\rm LS}$
and the direction to a galaxy.
We did not apply the described above correction to the galaxies residing
within the LS, those with $V_{\rm LG} < 600$~\kms\ and $|SGZ| < 1.5$~Mpc.

\begin{figure*}
 \centering
\includegraphics[angle=0,width=6.5cm,clip=]{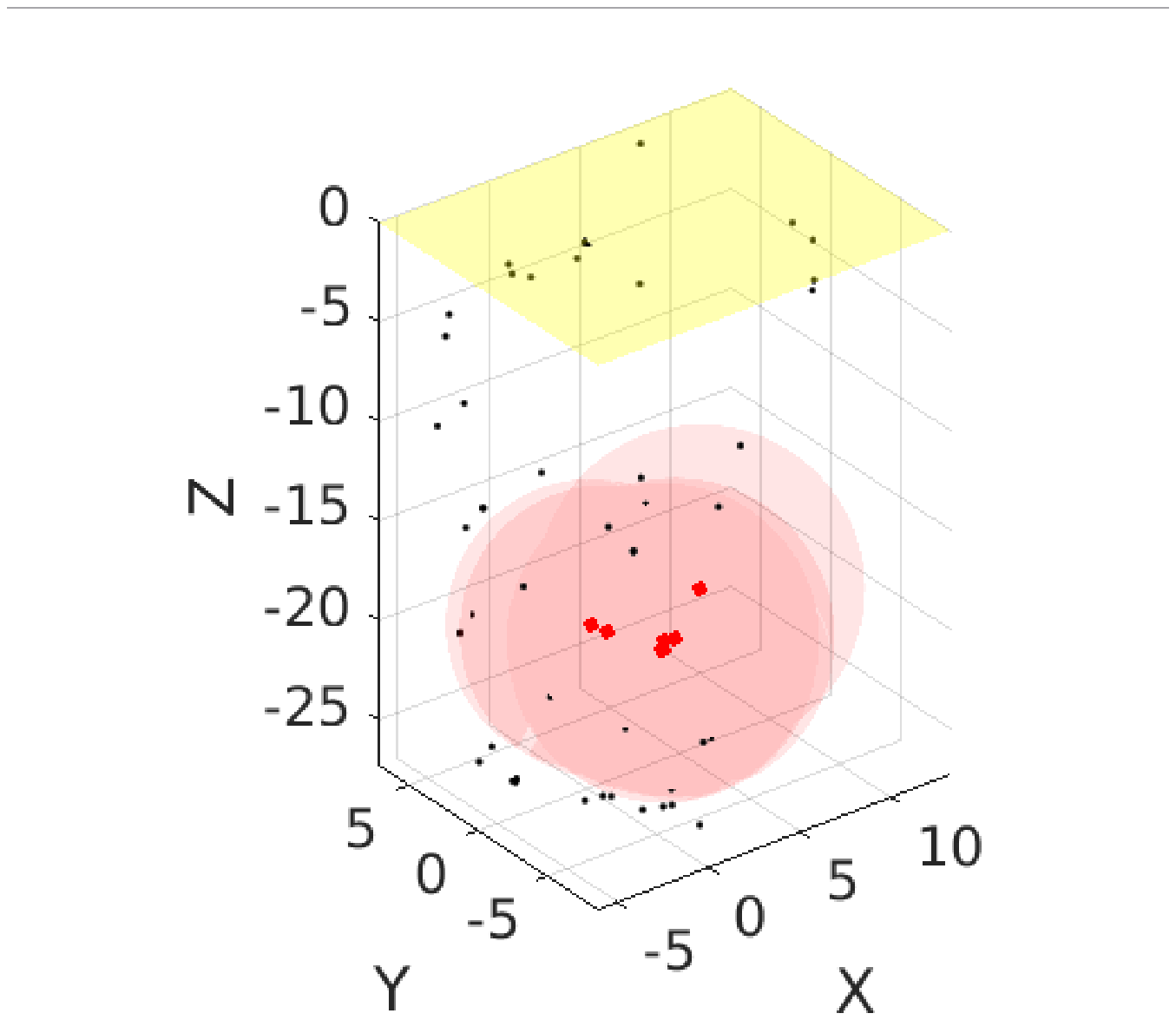}
\includegraphics[angle=0,width=6.5cm,clip=]{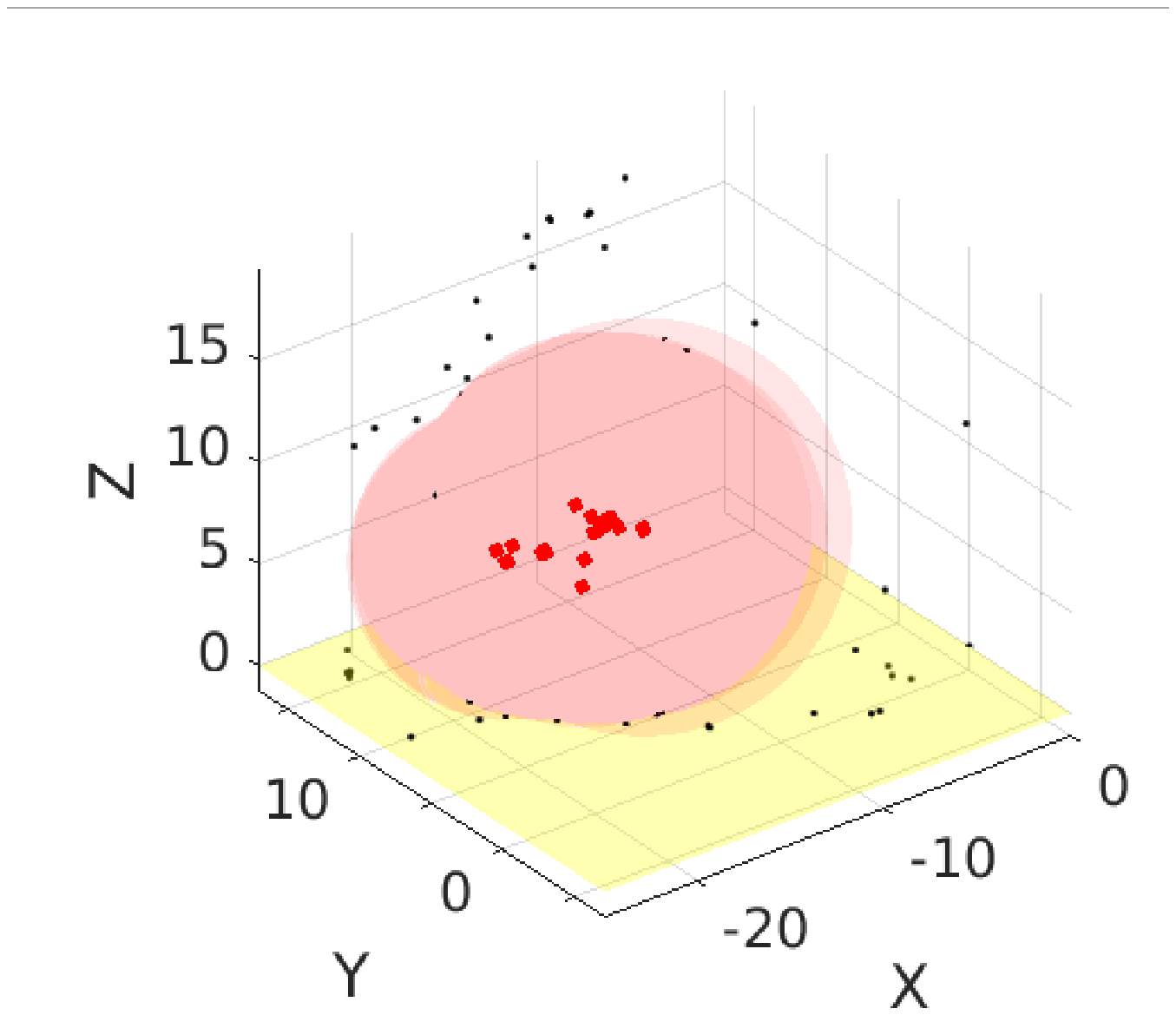}
  \caption{\label{fig:spheres}{\bf Left panel:} The 3D view in supergalactic
coordinates of empty spheres for void No.~11 (Monoceros, see
Sect.~\ref{ssec:nearby_voids}).
Black dots show positions of 'luminous' objects delineating voids. Red filled
circles show centers of individual empty spheres.
This sketch illustrates how a void is formed from the original lumped empty
spheres.
The position of the plane $SGZ=0$ is shown in aid to eye.
{\bf Right panel.} Similar view of empty spheres for void No.~19 (Lib).
  }
\end{figure*}

We used the known velocity-independent distance estimates in cases when
they came from the high precision methods based on the resolved stellar
photometry
(the tip of the red giant branch 'TRGB', Cepheids, supernova type Ia)
or via surface brightness fluctuations (SBF),
which in general give the accuracy of distances at 10 per cent level
\citep[e.g.][]{Tully09}.
When galaxies have several independent estimates with these methods, 
we combine them to derive the most robust value.

We select all galaxies within $D = V_{\rm cor}/H_0 < 25$~Mpc from us.
This gave us initially the sample of 7150 objects.
This number includes all LEDA's `galaxies' satisfying the above distance
limit. As one can see below,  358 of them (among residing in voids) appear
different kinds of wrong objects (see details in Sec.~\ref{ssec:wrong}),
which were cleaned from the final sample. The total number of galaxies with
the distance limit of $D = 25$~Mpc, as resulted from the search with
HyperLEDA, appeared after additional cleaning for wrong objects and parts
of galaxies to be 6792.

\renewcommand{\baselinestretch}{0.8}

\setcounter{qub}{0}

\begin{table*}
\begin{center}

\caption{\label{tab:voids} Main parameters of nearby voids}
\vspace{-0.3cm}

{\bf
\begin{tabular}{rlllccccccl} \hline \\[-0.2cm]
\multicolumn{1}{c}{\#}     &
\multicolumn{1}{c}{Void name}   &
\multicolumn{1}{c}{RA$_{\rm c}$} &
\multicolumn{1}{c}{Dec$_{\rm c}$}  &
\multicolumn{1}{c}{Dist$_{\rm c}$}  &
\multicolumn{1}{c}{Max.ext.}  &
\multicolumn{1}{l}{Nu.}  &
\multicolumn{1}{l}{Nu.}  &
\multicolumn{1}{l}{Tot.}  &
\multicolumn{1}{l}{Inner}  &
\multicolumn{1}{l}{Notes} \\
&
\multicolumn{1}{c}{  }  &
\multicolumn{1}{c}{hours}  &
\multicolumn{1}{c}{degr}  &
\multicolumn{1}{c}{Mpc }  &
\multicolumn{1}{c}{$\Delta X$,$\Delta Y$,$\Delta Z$}  &
\multicolumn{1}{c}{orig.}  &
\multicolumn{1}{c}{joined}  &
\multicolumn{1}{c}{void}  &
\multicolumn{1}{c}{void}  &
\multicolumn{1}{l}{  }  \\
&
\multicolumn{1}{c}{  }  &
\multicolumn{1}{c}{  }  &
\multicolumn{1}{c}{  }  &
\multicolumn{1}{c}{  }  &
\multicolumn{1}{c}{Mpc}  &
\multicolumn{1}{c}{sphere}  &
\multicolumn{1}{c}{sphere}  &
\multicolumn{1}{c}{gals}  &
\multicolumn{1}{c}{gals}  &
\multicolumn{1}{l}{  }  \\
&
\multicolumn{1}{c}{ (1) }  &                                   
\multicolumn{1}{c}{ (2) }  &                                   
\multicolumn{1}{c}{ (3) }  &                                   
\multicolumn{1}{c}{ (4) }  &                                   
\multicolumn{1}{c}{ (5) }  &
\multicolumn{1}{c}{ (6) }  &
\multicolumn{1}{c}{ (7) }  &
\multicolumn{1}{c}{ (8) }  &
\multicolumn{1}{c}{ (9) }  &
\multicolumn{1}{l}{ (10) }  \\
\\[-0.2cm] \hline \\[-0.2cm]
\qq  & Cas-And     &   00.7 & +53  &    19.0 &19,19,23&   13    &       5     & 19  &  15  &  \\      
\qq  & Tuc         &   00.9 & --64 &    11.2 &14,14,14&    5    &       1     & 56  &  44  & close \\ 
\qq  & Cet-Scu-Psc &   01.3 & --02 &    15.2 &33,17,29&   40    &      14     &108  &  85  &  \\      
\qq  & Pho         &   01.4 & --54 &    18.0 &17,19,18&   31    &       6     & 80  &  66  &  \\      
\qq  & Tau         &   03.8 & +17  &    18.8 &24,27,21&   17    &       7     & 53  &  46  & part at large extin. \\ 
\qq  & Per         &   04.0 & +52  &    19.7 &14,14,13&    2    &       1     &  4  &   3  & large extin. \\         
\qq  & Eri-Ori     &   05.1 & --07 &    18.5 &20,18,17&    5    &       2     & 41  &  30  &  \\                     
\qq  & Ori-Tau     &   05.4 & +15.2&    07.5 &18,14,14&   11    &       3     & 46  &  36  & close, part at large extin.\\
\qq  & Aur         &   05.8 & +38  &    13.5 &23,22,21&   14    &       4     & 36  &  34  & close, part at large extin.\\
\qq  & Lep         &   05.85 &--17 &    06.3 &13,13,13&    3    &       1     & 43  &  32  & close \\               
\qq  & Mon         &   06.4 &--07  &    20.2 &19,15,16&    9    &       3     &  7  &   2  & large extin. \\        
\qq  & Cnr-CMi-Hyd &   08.5 & +10  &    17.5 &29,25,23&   40    &      14     & 129 & 106  &  \\                    
\qq  & Vel         &   09.5 &--50  &    19.0 &20,27,21&   29    &       7     & 76  &  59  & half at large extin. \\
\qq  & Hyd         &   09.8 &--15.2&    19.2 &18,14,17&    8    &       3     & 22  &  15  &  \\                    
\qq  & Cen-Cir     &   14.4 &--65  &    21.0 &15,14,15&    4    &       1     & 20  &  14  & at large extin. \\     
\qq  & UMa         &   14.8 &+59   &    21.0 &15,16,13&    6    &       2     & 82  &  73  &  \\                    
\qq  & Vir-Boo     &   14.8 &+07   &    10.2 &14,14,15&    5    &       2     & 20  &  12  & close \\               
\qq  & Boo         &   15.3 & +27  &    19.4 &20,24,26&   11    &       4     & 49  &  33  &  \\                    
\qq  & Lib         &   15.4 &--26.5&    18.3 &19,21,24&   23    &       7     & 40  &  31  &  \\                    
\qq  & Her         &   16.6 &+20   &    13.5 &21,22,25&   28    &       8     & 118 &  97  & close \\               
\qq  & Oph-Sgr-Cap &   18.5 &--18  &    13.0 &37,30,35&   68    &      23     & 121 &  89  & incl. LV, part at large extin. \\
\qq  & Dra-Cep     &   20.4 &+71.1 &    13.9 &21,21,17&   16    &       5     & 50  &  44  & close, part at large extin. \\ 
\qq  & Cyg         &   20.6 &+36.3 &    19.3 &22,23,28&    7    &       3     & 20  &  18  & part at large extin. \\
\qq  & Pav-Oct     &   20.7 &--73.1&    15.7 &23,20,18&   19    &       5     & 36  &  30  & part at large extin. \\
\qq  & Aqu         &   22.7 &--02.5&    15.5 &20,22,24&   33    &       9     & 80  &  72  &  \\                    
\\[-0.25cm] \hline \\[-0.2cm]
\end{tabular}                                                                                                                                                                                                      
}
\end{center}
\end{table*}

The goal of the work is to construct an unbiased sample of voids on
the scale of 25~Mpc. To avoid the edge effects on the properties of
voids and its galaxy population, we extended the sample of `luminous'
galaxies up to the distance of $R = 28.0$~Mpc. This allowed us to
identify all empty spheres in the whole considered volume  up to
$R=25$~Mpc. The number of the `luminous' objects in our sample is 464,
of which 113 are the individual luminous galaxies and 351 are the known
aggregates around luminous galaxies.

On the first step, we generated the set of maximal empty spheres
inscribed into 3D distribution of the `luminous' galaxies and its groups
using Voronoi tessellation technique. The procedure for voids construction
begins with the selection of the largest empty sphere, which is the core
of a new void.
This void consists of all spheres whose centres lie within the half radius 
from the centre of the main sphere. On the next step, we find the next maximal
empty sphere and repeat the generation of a void around it.

We played with various $D_{\rm min}$ of empty spheres and the
distances to the outer boundary.
Our procedure selects at the first pass 445 empty spheres with radii from
11.0 to 6.0~Mpc gathered in 25 detached complexes. These complexes contain
from 2 to 68 original empty spheres.
However, all original empty spheres form clusters (in average of 3 spheres)
of very close spheres with the distances between their centres less than
1--2 Mpc as compared with the radii of the spheres of 6--11 Mpc.
If we join all empty sphere in clusters which we conditionally call the
'fat' sphere, the total number of 445 empty spheres reduces to 140 `fat'
empty spheres. These 'fat' spheres in turn are gathered in 25 detached
complexes which we identify as separate voids.
These voids and their more detailed description are presented in
Table~\ref{tab:voids} and Sec.~\ref{ssec:nearby_voids}.
In figure \ref{fig:spheres} we show 3D views of two voids (No.~11 and 19)
along with all empty spheres of which they are constructed (in supegalactic
coordinates SGX,SGY,SGZ). Small black dots show delineating luminous
galaxies/groups, while the filled red circles show positions of the lumped
empty sphere centers.

\section{Sample of galaxies in nearby voids}
\label{sec:voidgal}

When the nearby voids are defined as the aggregates of several/many
large empty spheres, we separate all galaxies which fall into interiors
of the individual empty spheres belonging to the defined voids.
This preliminary selection resulted in the list of 2075 entries.

Each galaxy falling into voids was checked to possible membership
of suits around 'luminous' border galaxies. The procedure followed the
methodology proposed by \citet{PaperI}.
Namely, we based on results
of \citet{Prada03} on the Dark Matter density profile and related
velocities of companions of massive galaxies at various radial distances.
For all separated 'void' galaxies with the projected distance to the
nearest luminous galaxy less than 1.2 Mpc, we examined the combination
of the projected distance and the difference in radial velocities of
a tentative host and a companion. If the difference was less than the
expected value as predicted in the model
of \citet{Prada03}, the galaxy was adopted as a companion and not
a void sample representative. Otherwise, it was considered as a void galaxy.
The threshold projected distance of 1.2~Mpc was selected close to the
radius of zero velocity of typical groups in the Local Volume
\citep{Karach2005}, and smaller than the distance to the nearby dwarf group
 NGC~3109 ($d \sim$1.3~Mpc), considered as a detached aggregate not bound
to the Local Group.
For example, galaxy SDSSJ092036.5+494031 is situated at the projected distance
of 0.33 Mpc from the luminous galaxy NGC2841 with $M_{\rm K}=-24.5$ and has
the relative radial velocity of $\delta V =$118~\kms. According to results
of \citet{Prada03}, this $\delta V$ is within the range found for bound
companions.

The redshift space distortions near the massive aggregates
such as galaxy clusters, lead to the well known  fingers-of-god pattern.
Thus, the cluster members can mimic the field object in the respective
sky areas. We took this effect into account using the clusterization
algorithm by \citet{LSCGroups}.   
Thereby, a void border object contains not only the bright
galaxy itself, but also all its satellites.
After this step, we also examined the remaining galaxies for their possible
membership of periphery in large known galaxy  aggregates
using the most recent information, namely, for the Virgo Cluster from
\citet{Virgo10, Virgo-front}, for UMa cluster complex \citep{UMa2013}, for
Leo Clouds \citep{Stierwalt09} and for Fornax cluster \citep{Fornax11}.
However, in case of complicated structures, a few interlopers could
remain in the sample.

These steps led to the exclusion of  361 galaxies from the initial
void sample. Also, during various verifications,
we excluded 312 various wrong entrees and 48 objects with poor quality
of redshifts. This issue is discussed in more detail
in Sec.~\ref{ssec:wrong}.

After the cleaning, the resulting sample of 1354 objects represents the
'expanded' void galaxy sample. It includes both galaxies:
a) situated substantially deep inside the voids (1088), and b) those falling
into transition outer zones of voids (266). The representatives
of this expanded sample are suitable for some of related studies, including
clustering of void galaxies and finding filaments and small scale structures
inside the studied voids.

\renewcommand{\baselinestretch}{0.8}

\setcounter{qub}{0}

\begin{table*}
\begin{center}

\caption{\label{tab:voidgal} Main parameters of galaxies in
the nearby voids}

\vspace{-0.3cm}

\scriptsize{\bf {
\begin{tabular}{rlcrrrrllllrl} \hline \\[-0.2cm]
\multicolumn{1}{c}{\#}     &
\multicolumn{1}{c}{Name}   &
\multicolumn{1}{c}{Coordinates (J2000)} &
\multicolumn{1}{c}{$V_{\rm hel} \pm \sigma_{\rm V}$}  &
\multicolumn{1}{c}{$V_{\rm LG}$}  &
\multicolumn{1}{c}{$V_{\rm dist}$}  &
\multicolumn{1}{c}{$B_{\rm tot}$}  &
\multicolumn{1}{c}{$A_{\rm B}$}  &
\multicolumn{1}{c}{$M_{\rm B}$}  &
\multicolumn{1}{c}{$M_{\rm K}$}  &
\multicolumn{1}{r}{$D_{\rm NN}$} &
\multicolumn{1}{l}{Void}         &
\multicolumn{1}{l}{Notes}       \\
&
\multicolumn{1}{c}{ (1) }  &                                   
\multicolumn{1}{c}{ (2) }  &                                   
\multicolumn{1}{c}{ (3) }  &                                   
\multicolumn{1}{c}{ (4) }  &                                   
\multicolumn{1}{c}{ (5) }  &
\multicolumn{1}{c}{ (6) }  &
\multicolumn{1}{r}{ (7) }  &
\multicolumn{1}{r}{ (8) }  &
\multicolumn{1}{r}{ (9) }  &
\multicolumn{1}{c}{ (10)}  &
\multicolumn{1}{c}{ (11)}  &
\multicolumn{1}{c}{ (12)}  \\
\\[-0.2cm] \hline \\[-0.2cm]
%
\qq& AGC102728 & J000021.4+310119 &  566   ~5 &  836 &  685 & 19.40 & 0.20 & -10.66 &   -  & 4.17  & 25 &   \\ %
\qq& PGC000083 & J000106.5+322241 &  542   ~8 &  815 &  667 & 18.04 & 0.22 & -11.99 &-14.72& 4.09  & 25 &   \\ %
\qq& ESO149-013& J000246.7-524618 & 1498   12 & 1418 & 1259 & 15.71 & 0.07 & -15.55 &-17.90& 4.70  & 04 &   \\ %
\qq& PGC000285 & J000406.3-572333 & 1889   64 & 1786 & 1640 & 16.55 & 0.08 & -15.29 &-17.63& 2.42  & 04 &   \\ %
\qq& PGC000389 & J000535.9-412856 & 1500   57 & 1475 & 1278 & 18.23 & 0.07 & -13.06 &    - & 3.26  & 03 & ++\\ %
\qq& ESO293-034& J000619.9-413000 & 1512   ~6 & 1487 & 1279 & 13.62 & 0.07 & -17.67 &-21.80& 3.28  & 03 & ++\\ %
\qq& PGC000482 & J000625.4-413024 & 1428   62 & 1402 & 1279 & 15.10 & 0.07 & -16.11 &-18.57& 3.28  & 03 & ++\\ %
\qq& AGC748778 & J000634.3+153039 &  258   ~5 &  486 &  463 & 18.91 & 0.28 & -10.38 &   -  & 4.38  & 25 &   \\ %
\qq& ESO293-035& J000651.6-415023 & 1552   ~7 & 1525 & 1280 & 16.77 & 0.06 & -14.51 &-16.86& 3.29  & 03 &   \\ %
\qq& AGC239031 & J000652.8-412530 & 1435   81 & 1410 & 1279 & 17.92 & 0.07 & -13.37 &-15.97& 3.30  & 03 &   \\ %
\\[-0.25cm] \hline \\[-0.2cm]
\multicolumn{13}{l}{Small part of Table~A1 is shown as an example. The entire Table~A1
is available in electronic form of the paper in the Appendix (On-line data).  } \\
\end{tabular}                                                                                                                                                                                                      
}
}
\end{center}
\end{table*}

In the context of the analysis of void galaxy evolution, by analogy with our
previous study of the galaxy sample in the nearby Lynx-Cancer void, we
introduce a more strict definition for an 'inner' void galaxy. The 'inner'
sample galaxies should be well isolated from their nearby luminous neighbours
that would warrant their secular evolution without significant disturbing
interaction from the luminous/massive neighbours. This, of course, does not
exclude possible interactions with smaller void galaxies. However, since the
number density of void galaxies is reduced in average by factor of 5--10,
the latter effect should be also much reduced with respect of similar
galaxies in more typical environments. For these 'inner' void sample galaxies,
we adopt the condition to be separated by no less than 2.0 Mpc from
the nearest 'luminous' galaxy, that is $D_{\rm NN} \geq $ 2.0 Mpc.

The remaining part of void galaxies with $D_{\rm NN} < $ 2.0 Mpc we
call the 'outer' void subsample. Their $D_{\rm NN}$ fall in the range of
0.5--1.99 Mpc, with the median $D_{\rm NN,median} = $ 1.64~Mpc.
In the future study of possible evolutionary differences of void
galaxies with respect to their counterparts in denser environments,
we need to exclude an uncontrolled effect of the bordering luminous
galaxies and/or an ambigious origin of 'outer' void galaxies due to
peculiar velocities and factor of uncertainty in their distances.
Therefore we do not use this subsample for galaxy evolution study
similar to our work on the Lynx-Cancer void sample.

The 'outer' void subsample can be important for more general studies,
not related to their evolution issues. In particular, in the study of
void substructures, one uses all void population since we expect the
continuity of void filaments all the way from the central parts to
its boundary. One should also
understand that the threshold value of $D_{\rm NN} =$2.0 Mpc is a
somewhat conditional choice. Due to the distance uncertainties, some
galaxies with $D_{\rm NN}$ a little smaller than 2.0~Mpc may be in fact
the representatives of 'inner' sample and thus, can (and do) share
properties of this sample galaxies.

\section[]{RESULTS}
\label{sec:results}

\subsection{Identified nearby voids and their properties}
\label{ssec:nearby_voids}

In this section we present the identified nearby voids,
their main parameters, regions of the sky, and relation
to the nearby voids described in earlier studies.
The main information on nearby voids is summarized in
Table~\ref{tab:voids}.
Besides the number of each void in our 'Nearby Void Sample', we also present:
in Col.~1 - a void name related to the respective constellation where the main
part of void is projected.  Columns~2, 3 and 4 present the approximate
positions of void centres (respectively, RA, Declination and the distance
in Mpc). In Col.~5 we give the approximate maximum extent of each void along
the supergalactic X, Y, Z.
Col.~6 and 7 give the numbers of original empty spheres and the number of
lumped 'fat' spheres in each void. Col.~8 gives the total number of galaxies
in a void, while in Col.~9 we present the respective number of 'inner' void
galaxy subsample.
In Col.~10 additional notes are given.

The largest void complex, as one could expect, is associated with
the region of the Local Void and its surroundings. In our notation
this is the complex of empty spheres No.~21 in Ophiuchus - Sagittarius -
Capricornus. Its original 68 empty spheres
with radii $R \sim 8$--10~Mpc are clustered near $\textrm{R.A.}\approx 18.5^h$
and $\textrm{Dec.}\approx-18\degr$, with distances to their centres of
$D \sim$12--17~Mpc.
Approximately at the opposite side with respect of the Local Sheet
there is another large void, No~12, in Cancer - Canis Minoris - Hydra.
As mentioned in Introduction, we already studied a part of this void under
the name Lynx-Cancer void.

As one can see in Table~\ref{tab:voids}, the median diameters of nearby voids
are $\sim$19~Mpc, with the full range of 13 to more than 30~Mpc. From this,
the typical volume of separated voids is of $\sim$3500~Mpc$^3$, with the full
range from $\sim$1000~Mpc$^3$ to $\sim$19000~Mpc$^3$. It is interesting to
compare the selected nearby voids with the minivoids found by \citet{TK06}
within the volume of $R < 7.5$~Mpc. Their largest
voids completely devoid of known (at that time) galaxies have volume of
$\sim$60--440~Mpc$^3$ and probably comprise the fine substructure of larger
voids defined in this paper.
Possible examples of such fine structure are the fragments of void filaments
suggested in \citet{Beygu13} and \citet{UGC3672A}.

\begin{figure*}
 \centering
\includegraphics[angle=0,width=14cm,clip=]{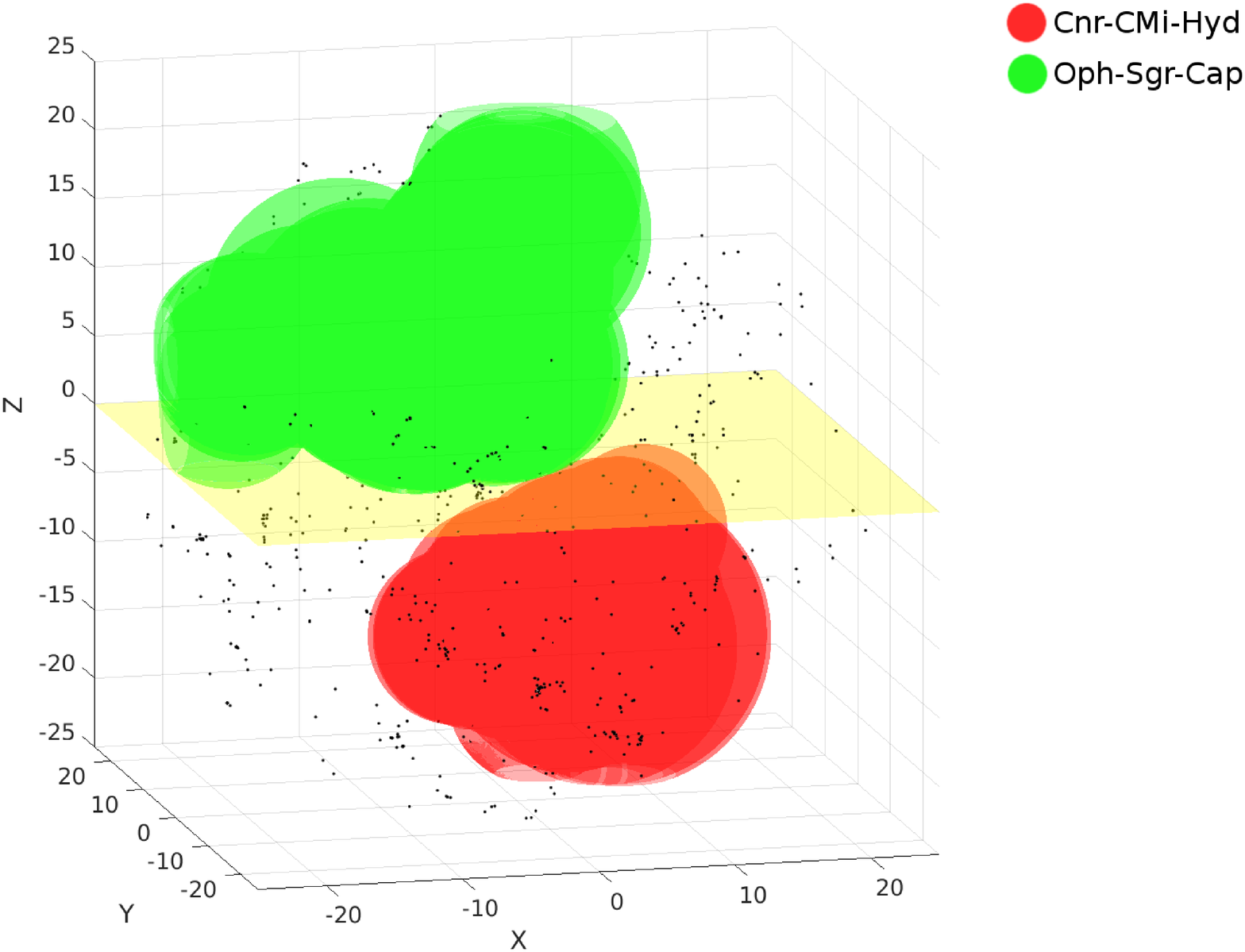}
\includegraphics[angle=0,width=14cm,clip=]{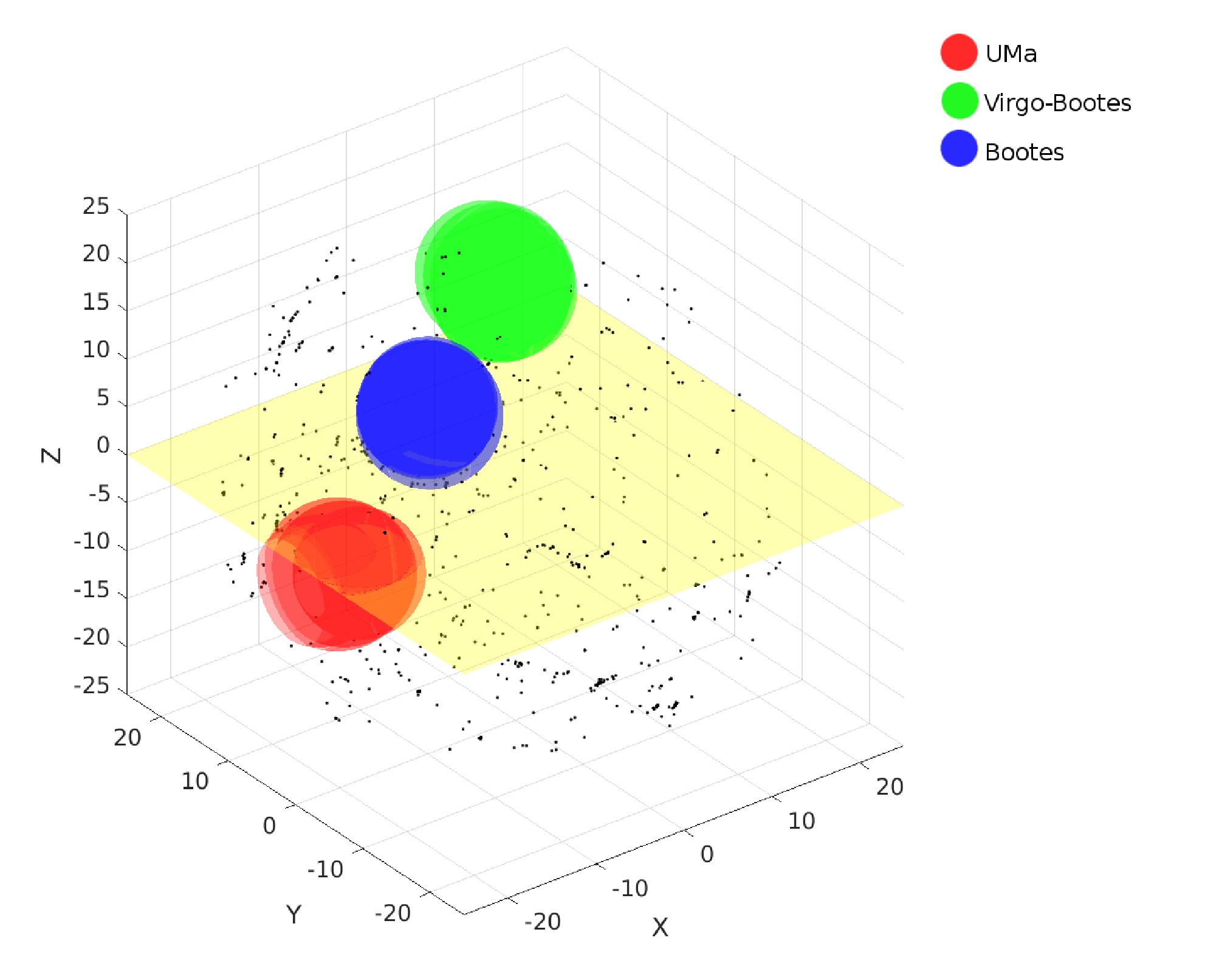}
  \caption{\label{fig:voids.3D}{\bf Top panel:} The 3D view in supergalactic
coordinates of the two largest nearby voids described in
Sect.~\ref{ssec:nearby_voids}, Cancer--Canis-Minor--Hydra and
Ophiuhus--Sagittarius--Capricornus (includes the near part of the
Local Void and its environs). The position of the plane $SGZ=0$ is shown
in aid to eye.
{\bf Bottom panel.} Similar view of three smaller voids in
Ursa Major (red), Virgo--Bootes (green) and Bootes (blue).
  }
\end{figure*}

In Appendix we present 3D figures for all individual voids as positioned
with respect of the Local Sheet. As an example, in the top panel of
Fig.~\ref{fig:voids.3D}, we show the 3D view of two the largest voids while
in the bottom panel the similar 3D view of several more typical nearby
voids is presented (see Figure legends).

\subsection{Nearby voids galaxy sample}
\label{ssec:void_gals}

In this section we present the lists of galaxies which reside in
the nearby voids. Altogether the general sample includes 1354 objects.
As was explained above, we divide them into two subsamples and present
them in separate tables.
The main 'inner' sample with the distances to the nearest luminous bording
galaxy of $D_{\rm NN} \geq 2.0$ Mpc includes 1088 galaxies. The remaining
subsample of 'outer' (periphery) void galaxies with $D_{\rm NN} < 2.0$ Mpc
includes 266 objects.

\begin{figure}
 \centering
\includegraphics[angle=-90,width=8.0cm,clip=]{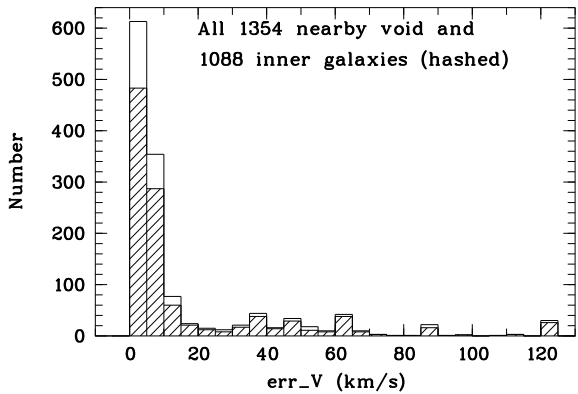}
  \caption{\label{fig:histo_errV} Distribution of the adopted velocity error
$\sigma_{V}$ (in \kms)  for the total 1354 and 1088 inner (hashed) void
galaxies. }
\end{figure}

The main known global parameters of void galaxies are adopted from LEDA
and NED. If some of the database parameters were found to be wrong, we used
a corrected/improved value estimated by us or taken from the recent papers.
In Table~\ref{tab:voidgal}   we give the first 12 strings of Table~A1 from
Appendix (On-line materials) in order to show its structure. Table~A1 presents
the list of all 1088 'inner' void galaxies with their main observational
parameters. It includes the columns. 1. Name (as in LEDA). 2. Coordinates on
the epoch J2000. 3. $V_{\rm hel}$ and its error $\sigma_{\rm V}$ (in \kms).
The mean $\sigma_{\rm V}$ = 17~\kms, with only a few per cent data with
$\sigma_{\rm V}$ above 70~\kms\ (see histogram in Fig.~\ref{fig:histo_errV}).
4. Corresponding velocity in the Local Group system, $V_{\rm LG}$,
5. Adopted velocity $V_{\rm dist}$, which is defined as
$D \times H_{\mathrm 0}$, where $H_{\mathrm 0} = $73~\kms~Mpc$^{-1}$ is
the adopted Hubble constant, and $D$ is distance in Mpc. For galaxies with
known independent distance estimates via the high accuracy methods (TRGB,
Cepheids, SNIa, etc), we adopt the respective value. For galaxies without
such data residing in the Local Sheet (where the peciliar velocity correction
is not applicable), $V_{\rm dist}$ is adopted to be $V_{\rm LG}$. For the
remaining galaxies outside the Local Sheet, we use for $V_{\rm dist}$ the
model of \citet{Tully2008}, with $V_{\rm dist} = V_{\rm LG} + \Delta V$,
where the peculiar velocity $\Delta V$ is a correction depending on galaxy
position. It takes into account the overall motion of the Local Sheet and
the Local Group from the Local Void and to the direction of Virgo Cluster.
Its maximal value is 357~\kms. See more detailes in Sec.~\ref{ssec:procedure}.
In column 6 we present the total $B_{\rm tot}$ magnitude, taken either from
LEDA, or, if absent, in NED. For part of the sample galaxies within
the SDSS footprint, without $B$ in LEDA and NED databases, we produced our
own estimate of $B_{\rm tot}$ based on the SDSS model $g,r$ galaxy magnitudes
via their transform to $B$-band according to formula from \citet{Lupton05}.
7. Adopted the Milky Way extinction in $B$-band, $A_{\rm B}$ (in magnitudes),
according to \citet{SF2011}.
8. Absolute blue magnitude $M_{\rm B}$ derived from the galaxy distance
($V_{\rm dist}$/73), $B_{\rm tot}$ and $A_{\rm B}$ in columns 5, 6 and 7.
In column 9 we give the absolute magnitude in $K$-band $M_{\rm K}$ as
presented in LEDA.
The latter is mostly based on the Two-micron survey data for brighter
objects, and on $M_{\rm B}$ and the mean colour $B-K$ for the
respective morphological type as found by \citet{UNGC13} for
fainter galaxies.
10. The distance $D_{\rm NN}$ (in Mpc) from a void galaxy to the nearest
luminous neighbour. 11. Number of a Nearby Void (in terms of
Table~\ref{tab:voids}) to which a void galaxy belongs.

The similar Table~A2 in Appendix presents  266 nearby void galaxies
residing in 'outer' parts of voids.

\section[]{DISCUSSION}
\label{sec:dis}

\subsection{Statistical data on void galaxies}

In total, within the sphere with $R = 25$~Mpc we identify 1354 galaxies
falling to the empty spheres belonging to 25 nearby voids
presented in Sect.~\ref{ssec:nearby_voids}.  Of them, 266 objects
have distances to the nearest luminous neighbours delineating the voids
($D_{\rm NN}$) less than 2.0~Mpc, that is they populate the outer void
regions. We call these galaxies, residing at the void periphery, the 'outer'
void objects. Due to various cosmological time-scale processes, on the one
hand, and due to the observational uncertainties in their distances, on the
other hand, a larger or smaller part of them can appear the objects from the
denser surroundings of voids. Therefore, if for the study of void galaxy
evolution we include this group, this can increase the scatter and shift the
mean and median of the clean sample parameters. Their median
$D_{\rm NN}$ = 1.65~Mpc, and
only 18 of these 266 galaxies have $D_{\rm NN}$ in the range 0.46--1.0 Mpc.

\begin{figure}
 \centering
\includegraphics[angle=-90,width=7.5cm,clip=]{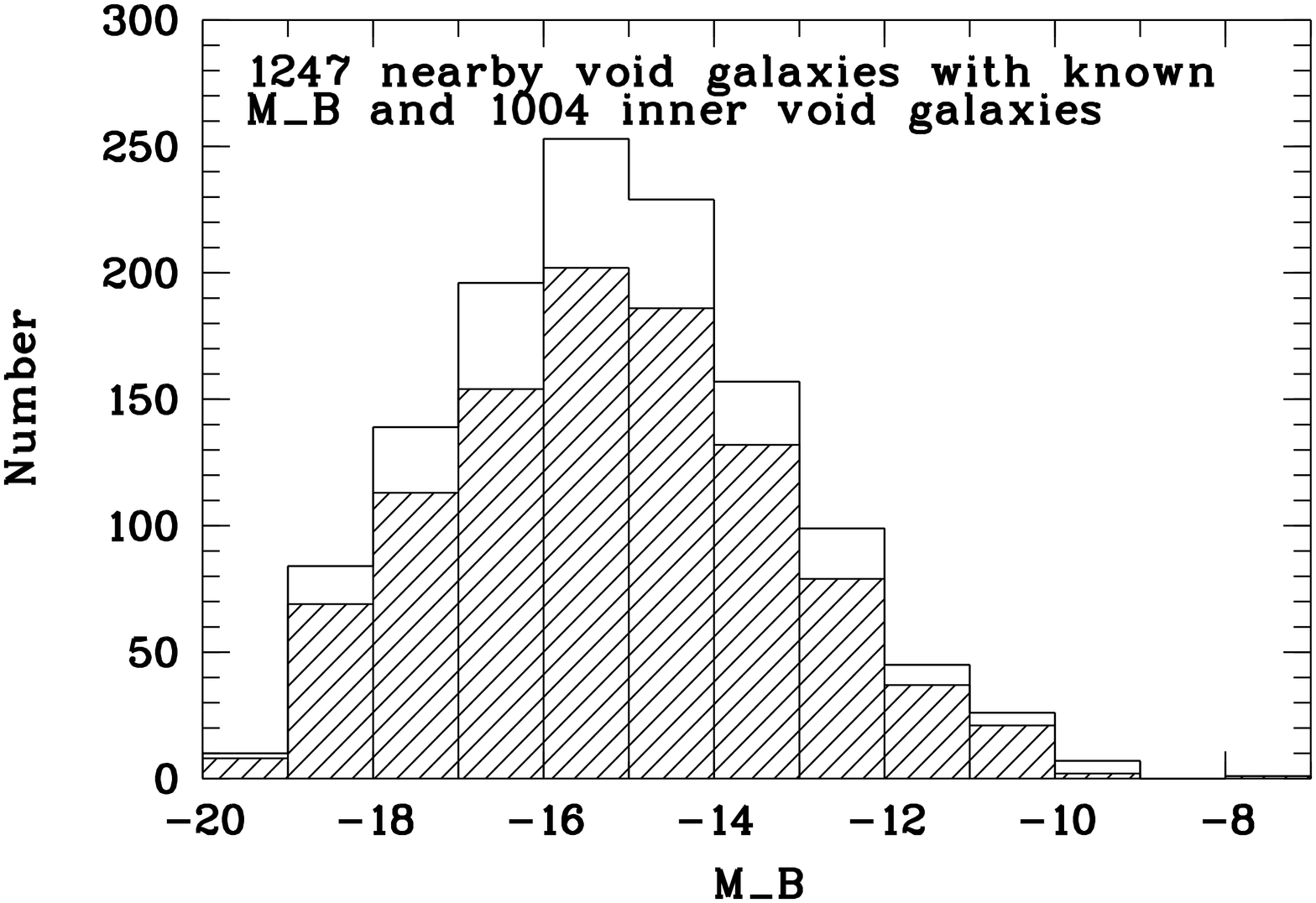}
\includegraphics[angle=-90,width=7.5cm,clip=]{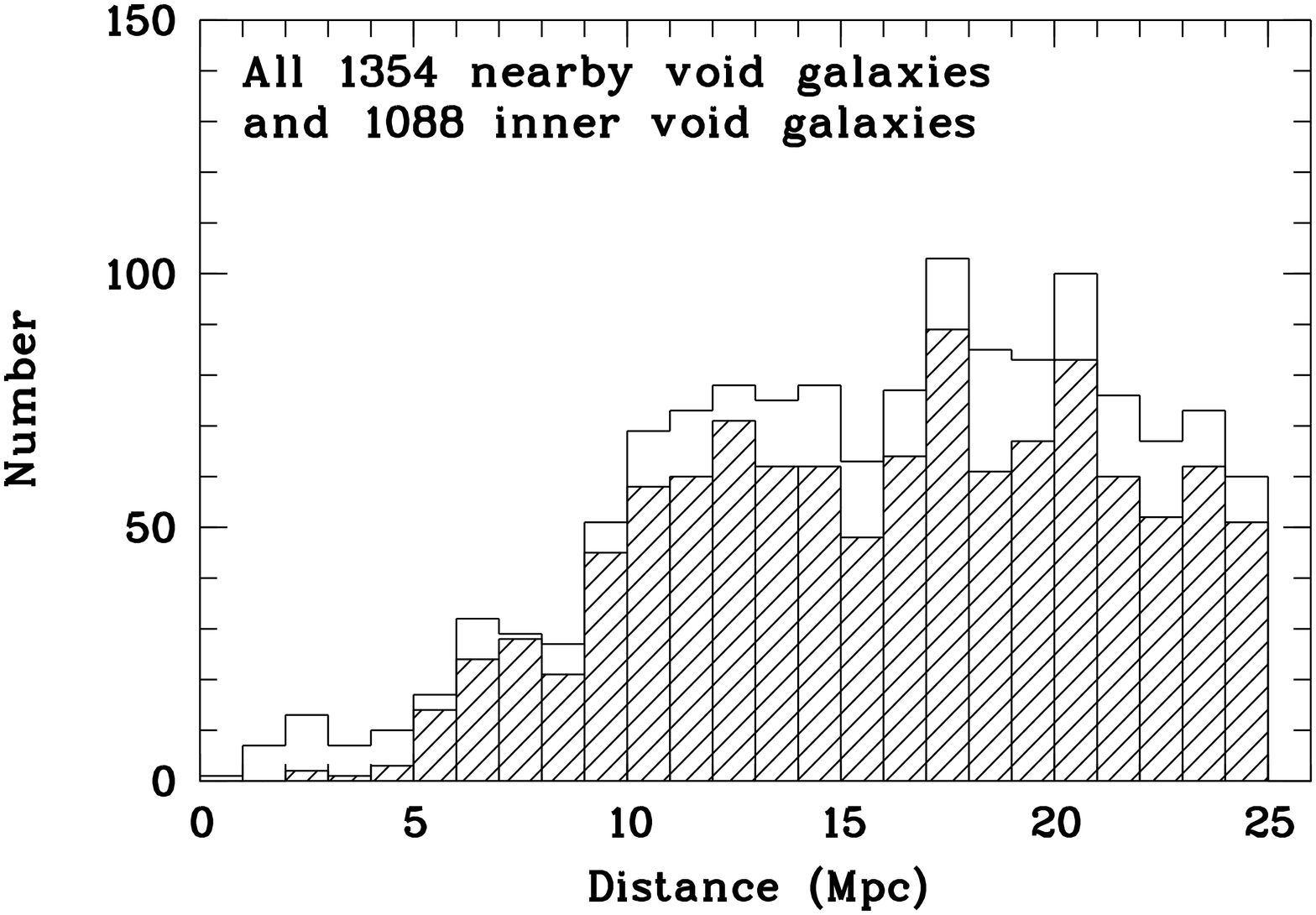}
\includegraphics[angle=-90,width=7.5cm,clip=]{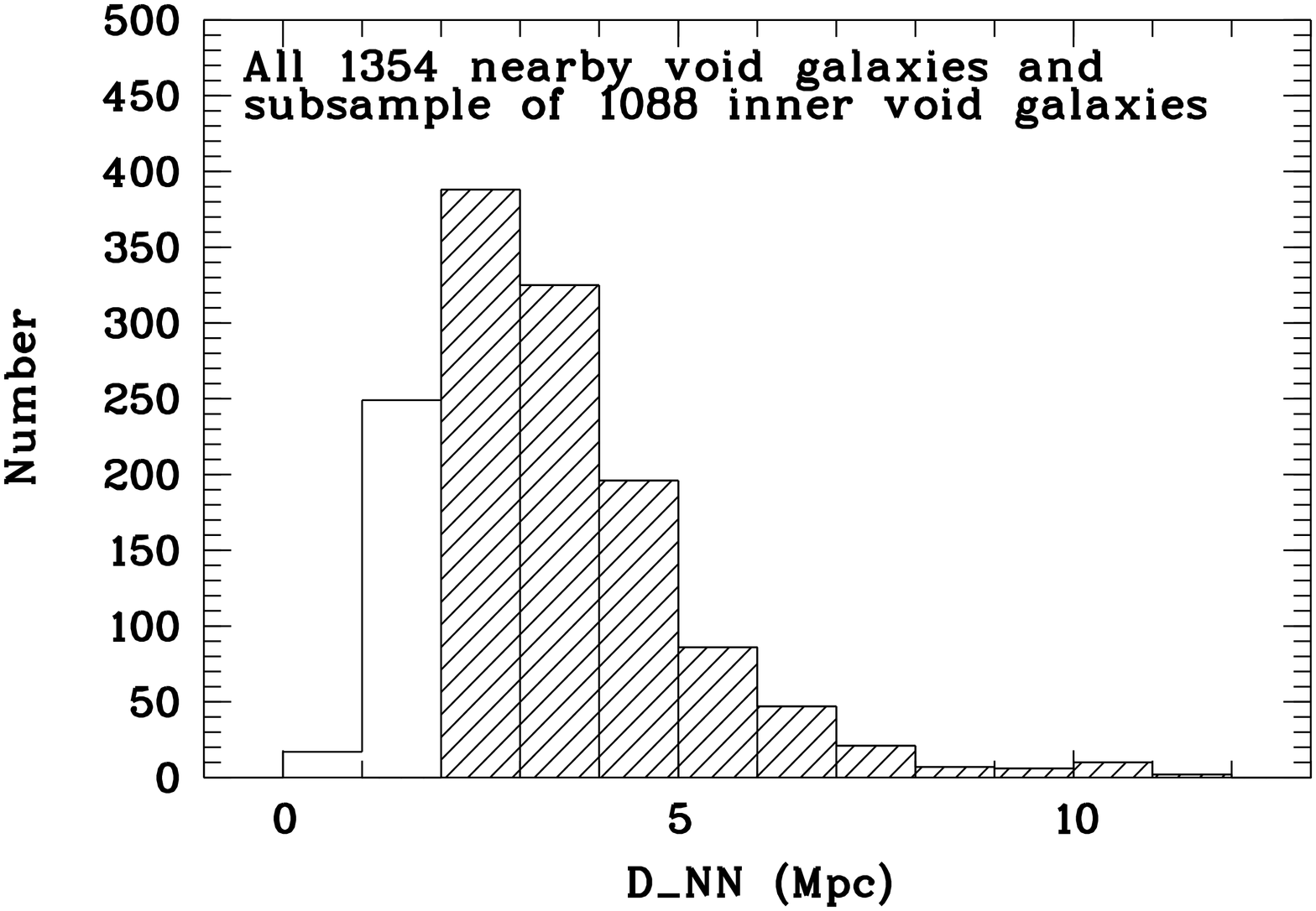}
  \caption{\label{fig:histo_NV}
{\bf Top panel:} Distribution of all 1247 and 1004 'inner' (shown by hashing)
void sample on $M_{\rm B}$ (for those with known $B_{\rm tot}$),
{\bf Middle panel:}  similar distibution on Distance for all 1354 and 1088
inner (hashed) void galaxies,
{\bf Bottom panel.} similar distribution on $D_{NN}$, the distance to the
nearest luminous object, for all 1354 and 1088 inner void galaxies. }
\end{figure}

The remaining part of 1088 galaxies with $D_{\rm NN} \geq $ 2.0~Mpc we call
conditionally the
'inner' void galaxies. They can be considered as genuine/native void objects
which formed and/or evolved within the low density environments. The adopted
threshold value of $D_{\rm NN}$ is an arbitrary, but according
to our previous work on the nearby Lynx-Cancer void, this allows us to
exclude efficiently various 'interloper' objects while adressing the
issue of possible peculiarities in the void galaxy evolution.
The median value of $D_{\rm NN}$ for the inner subsample is 3.43~Mpc,
with the full range of 2.0--11.5~Mpc. As one can see in
Fig.~\ref{fig:histo_NV}
(bottom panel), the $D_{\rm NN}$ distribution of inner subsample is skewed to
smaller distances, with 393 objects falling within the range of 2.0--3.0
Mpc. Of them, 204 inner galaxies have $D_{\rm NN}$ between 2.0 and 2.5 Mpc.
This is a reflection of the known concentration of void population to their
borders, derived both observationally and in simulations (see, e.g.,
\citet{Lindner96} and \citet{Gottlober03}).

When we study the spatial distribution and clustering
of void galaxies, the whole sample of void galaxies can be examined,
since substructures of the void galaxy distribution should be more
or less smoothly connected to those in the surrounding denser environments.

\begin{figure*}
 \centering
\includegraphics[angle=-90,width=8.0cm,clip=]{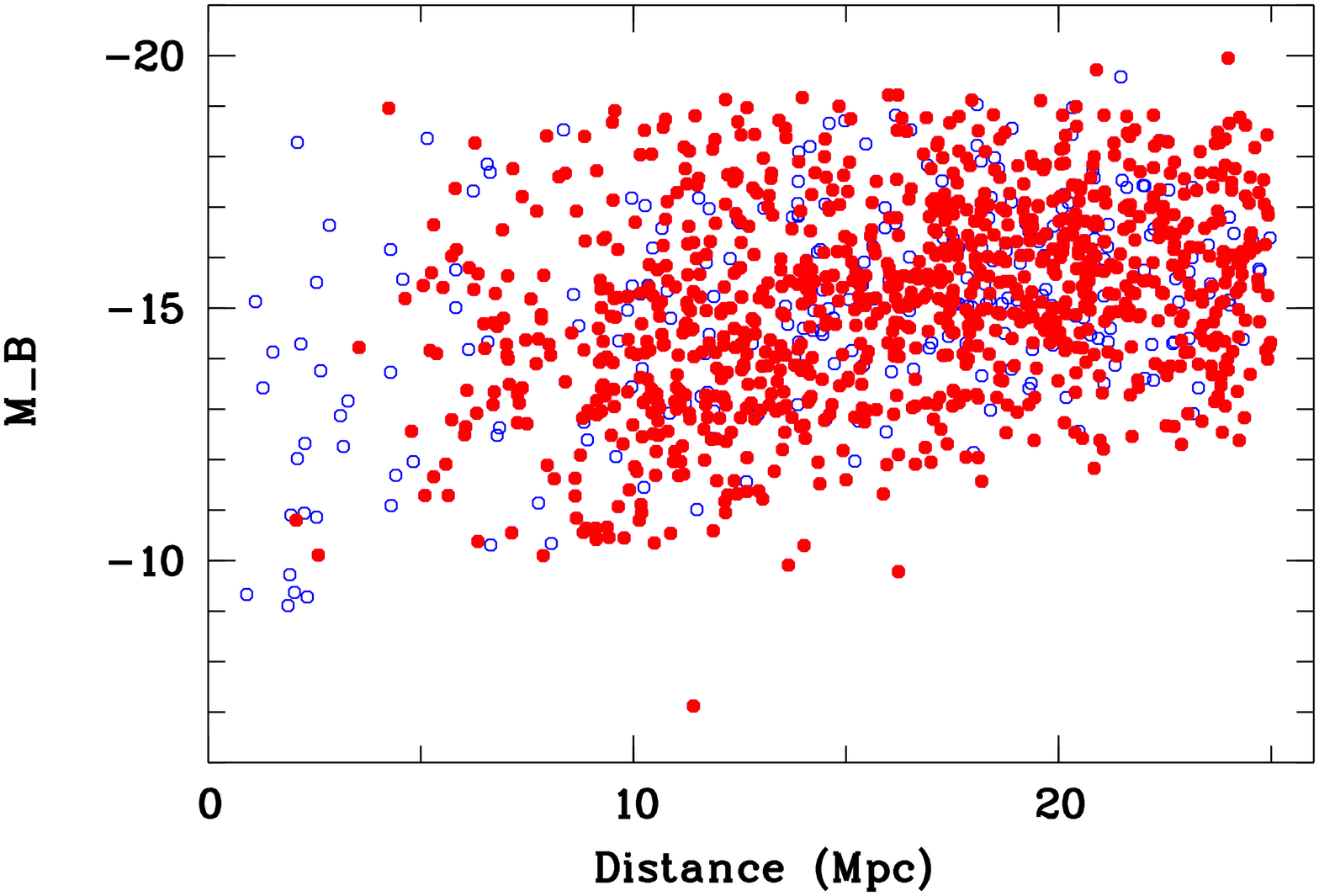}
\includegraphics[angle=-90,width=8.0cm,clip=]{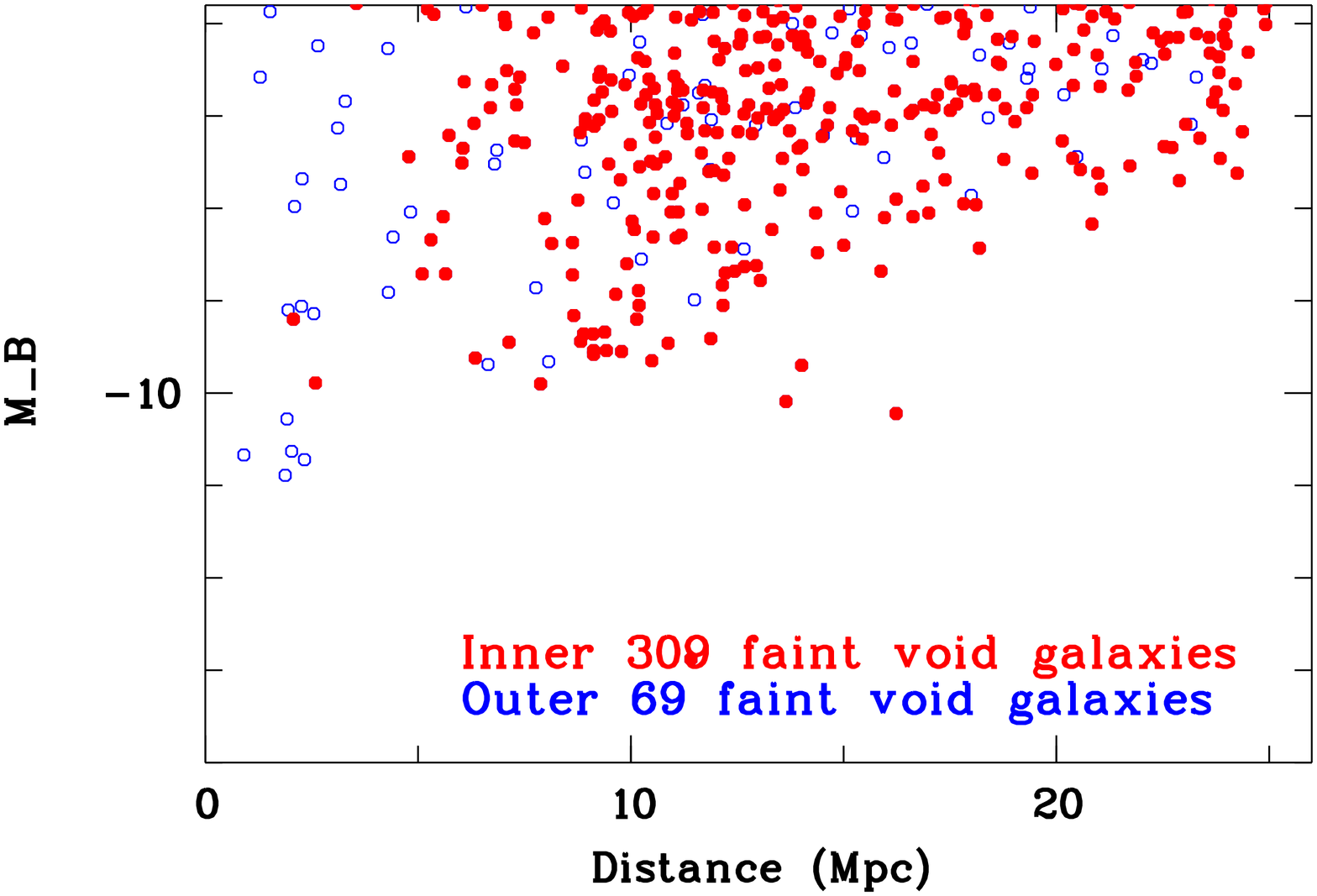}
  \caption{\label{fig:MBvsDist}
{\bf Left-hand panel:} Relation between the absolute magnitude $M_{\rm B}$ and
distance for all 1246 nearby void galaxies with known $B_{\rm tot}$. Red
filled octogons show 1004 'inner' void galaxies, blue open octogons -- 242
'outer' void galaxies.
{\bf Right-hand panel.} Close-up of this relation for the faintest 378
void galaxies ($M_{\rm B} > -14.2^{m}$) with known $B_{\rm tot}$.
309 'inner' void faint galaxies are shown by red filled octogons,
while 69 'outer' void faint galaxies - by blue open octogons.}
\end{figure*}

It is of interest, how much the effect of observational selection defines
the content of intrinsically faint dwarfs in the nearby voids. In
Fig.~\ref{fig:MBvsDist} we plot the relation of the void galaxies
$M_{\rm B}$ versus their distances.
In the left panel all 1246 nearby void galaxies are shown with known
$B_{\rm tot}$, of which red filled octogons show 1004 'inner' void
galaxies, while blue open octogons - 242 'outer' void galaxies.
In the right panel we show the similar plot, but with close-up of the
region for the faintest 378 void galaxies (conditionally,
$M_{\rm B} > -14.2^{m}$) with known $B_{\rm tot}$. Similary,
309 inner void faint galaxies are shown by red filled octogons,
while 69 outer void faint galaxies - by blue open octogons.

The lower boundary on these plots roughly corresponds (with a few
fainter exceptions) to the limiting apparent $B_{\rm tot} \sim 19.5$ mag.
The latter value is fainter than the typical limit for the main SDSS
spectroscopic survey $B \sim 18.5$ mag, as derived for typical dIs colours
$g-r$ and the adopted SDSS spectroscopic survey threshold of
$r \sim 17.8$ mag. This partly is caused by the use of \HI\ radial
velocities of the ALFALFA faint sources identified with the faint optical
galaxies with unknown redshifts and also by the dedicated search for
fainter companions of the known void dwarfs. For the Southern hemisphere,
the majority of the faintest galaxies come from the 2dF Galaxy Redshift
Survey (2dFGRS) \citep{2dFGRS}, where the limiting apparent
$B_{\rm tot}$ is close to 19.5 mag.

\subsection{Preliminary morphology classification of void galaxies}
\label{ssec:morpho}

The great majority of void galaxies comprise late-type objects
of Sd-Sdm at bright part of the luminosity function to various
dIs at the faint end (e.g., \citet{PaperI,Hoyle15,Beygu17}).
Below we attempt to summarize the available
information on morphological types of galaxies in the Nearby Voids.
There is no data for 271 objects, in HyperLEDA, NED or in the
literature. For 241 of these galaxies, we performed our own coarse
morphological classification based on the available CCD images of
the SDSS and PanSTARRS1 (PS1) surveys \citep{DR7,SDSS-IV,PanSTARRS1}.
In the sky regions outside the footprints of the above surveys, we
used the scanned digitized photographic images from the DSS. We can
not classify the remaining 30 objects, mainly due to the large
extinction in the Milky Way. These objects are discovered via the
blind \HI-line surveys of ZOA and as such most probably belong to
the late-type low mass galaxies. In the further statistics we count
them conditionally as late spirals.

With this reservation, the whole sample of 1354 nearby void galaxies
includes 95 objects (7.0 per cent) Early Type Galaxies (ETGs): E/dE
or E-S0;
46 (3.4 per cent) lenticulars (S0-S0a); 371 objects (27.4 per cent)
massive spirals (Sa-Sc and their subtypes); 192 (14.2 per cent) late
spirals (Sd-Sm), including the mentioned above 30 obscured objects;
585 (43.2 per cent) various irregular and dwarf irregular galaxies
(I/dI). Finally, we identify 16 blue compact galaxies (BCG) or
BCG-like objects (1.2 per cent), which we count separately.
Disc and irregular type galaxy population in voids is well
known and is studied in many papers cited above during at least the
last decade.

The existence in voids of 95 ETGs is very exciting since this gives
us an opportunity to study their formation and evolution in the low
density environment. Their real proportion still needs in careful
checks since a part of them can represent unrecognized cases of
projection of a Milky Way foreground star (with the radial velocity
of hundreds to a thousand \kms) onto the central part of a faint distant
galaxy. In the current version of the void galaxy sample, based on
the unbiased selection among all HyperLEDA objects classified as
galaxies, the fraction of ETGs does not exceed 0.07.
This amount is consistent with that found for the Void Galaxy Survey
(VGS), namely, 3 ETGs of 60 galaxies in the whole void sample
\citep{Kreckel12}.

Similary, the fraction of ETGs of $\sim$0.05 among a half thousand
nearby isolated galaxies (LOG sample) was found by \citet{LOG2011}.
These numbers hint that the fraction of ETGs outside groups and
clusters weakly depends on the degree of isolation. It will be the
task of the dedicated study to clear up the true nature of this type
void galaxy sample members and their origin in the void environment.

\begin{figure}
 \centering
\includegraphics[angle=-90,width=7.5cm,clip=]{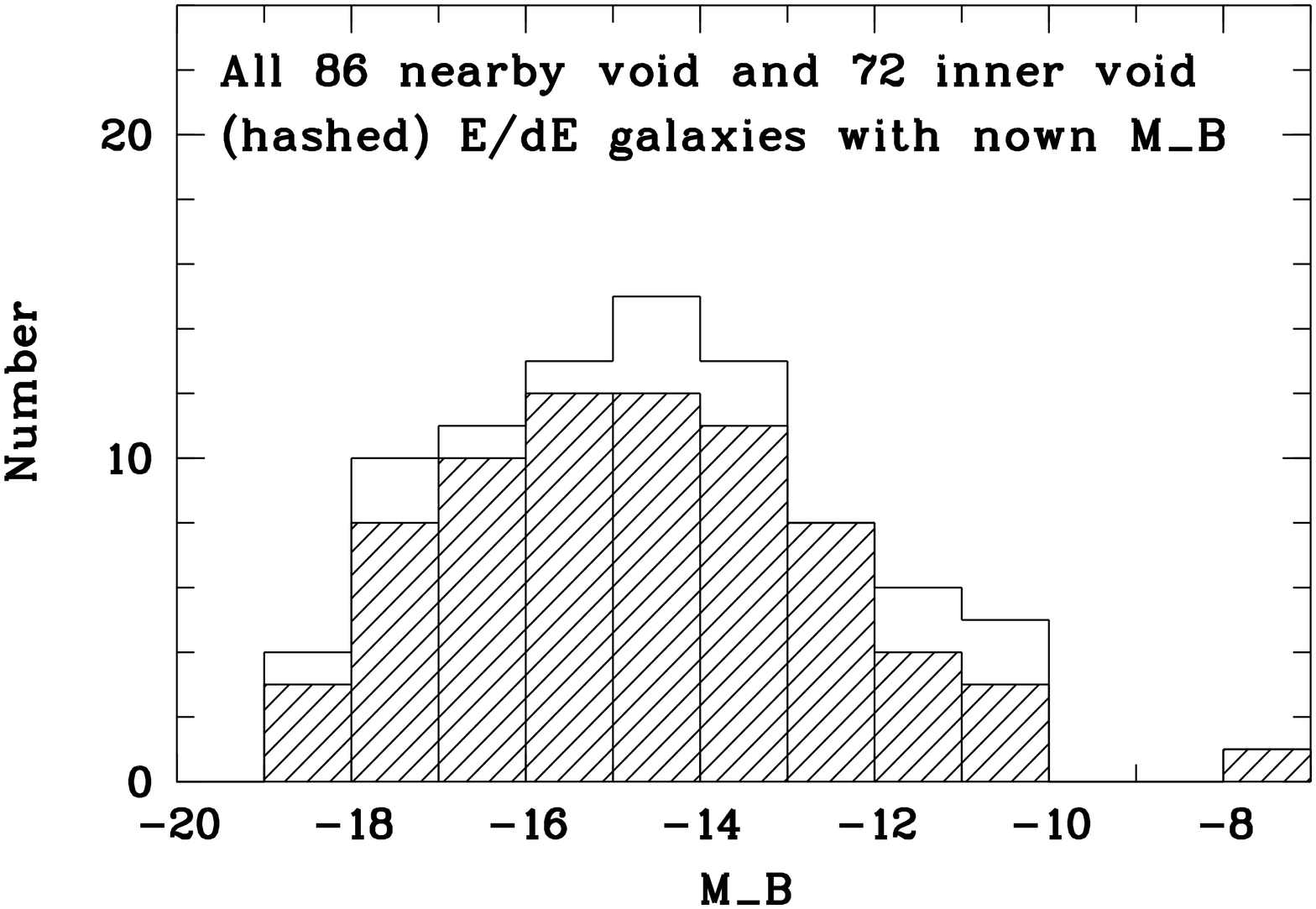}
\includegraphics[angle=-90,width=7.5cm,clip=]{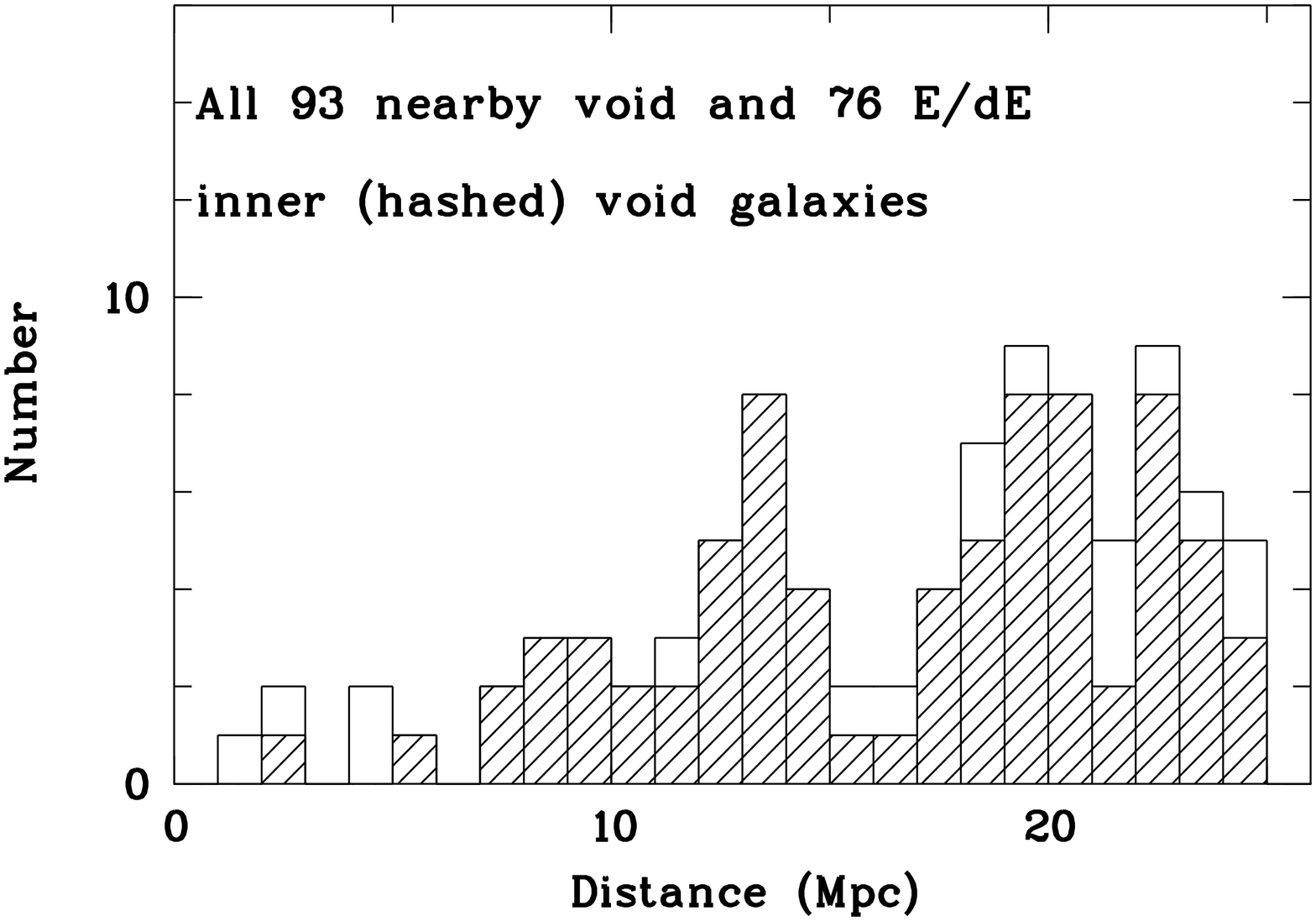}
\includegraphics[angle=-90,width=7.5cm,clip=]{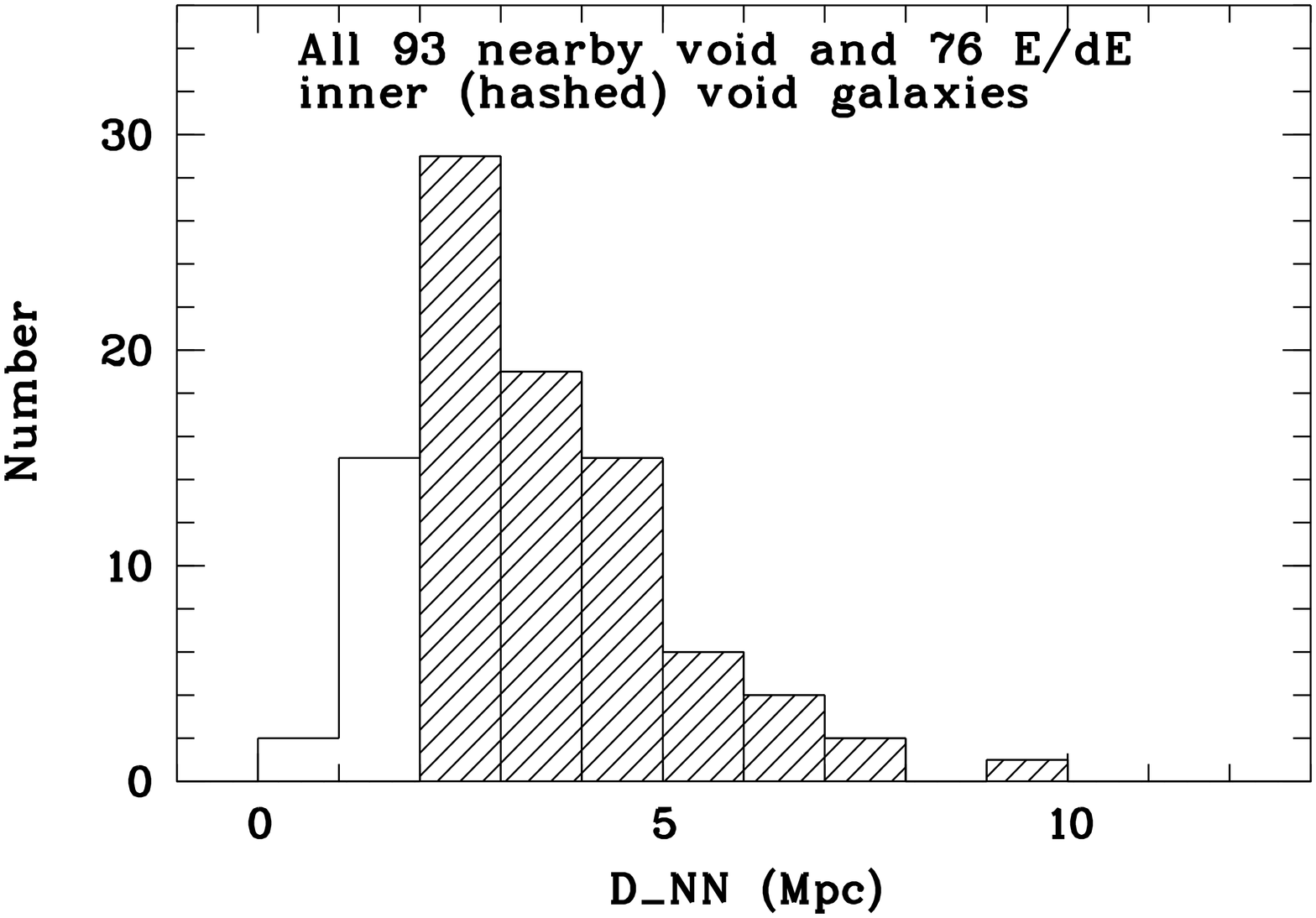}
  \caption{\label{fig:histo_ETG}
{\bf Top panel:} Distribution of all 86 and 72 'inner' (shown by hashing)
Early Type galaxies (ETG)  on $M_{\rm B}$ (for those with known $B_{\rm tot}$),
{\bf Middle panel:}  similar distibution on Distance for all 93 and 76
inner (hashed) void ETG,
{\bf Bottom panel.} similar distribution on $D_{NN}$, the distance to the
nearest luminous object, for all 93 and 76 inner void ETG. }
\end{figure}

We mention that the very rare cases of field isolated dE/dSph (like the
galaxy KKs~3) are found recently by \citet{LeoSpur15,KKs3}
at $\sim$2.1~Mpc from the Local Group. Their Tip of RGB (TRGB) distances
are based on the HST images. Probably dEs from the presented Nearby
Void galaxy sample have the common origin and evolutionary paths with KKs~3.
About 25 per cent of the early-type galaxies from the current Nearby Void
sample are detected in \HI-line in the frame of several known \HI-surveys.
This indicates their atypical evolutionary properties. In this context,
it is  worth to  mention the recent results of \HI\ survey of a small group
of isolated early-type galaxies (ETG) by \citet{Ashley17}, who indicate
their unusual properties with respect of the similar ETGs in dense
(groups and cluster) environments. The earlier work on \HI\ properties
of ETGs was presented by \citet{Grossi09} based on the ALFALFA data.
Some of them also reside in voids and show differences in their
properties in comparison to the typical ETGs.

\subsection{Void galaxies in and near the Zone of Avoidance}

As was mentioned in the Introduction and in Sec.~\ref{ssec:procedure}, thanks
to the definition of luminous galaxy sample delineating voids via their
$K$-band luminosity, we probe the low-density regions through the whole sky,
without the artificial break of structures near the Milky Way equator. Parts
of the identified voids, hence, cover the Zone of Avoidance (ZOA), the sky
region near the galactic equator (roughly, $|b| < 15$\degr). The related
amount of dust extinction in $B$-band, $A_{\rm B}$, varies for void galaxies
roughly from one to several (and in a few cases ten and more) magnitudes.
Specifically, we identify for the inner void galaxy sample 126 galaxies with
$A_{\rm B} > 1.0$ mag.

The luminous galaxies till the distance of 28 Mpc, used to define the nearby
voids, all are visible within the ZOA. However, the optical detection of
smaller galaxies and dwarfs is affected by both the above substantial
extinction and by the crowded stellar environments. Therefore, the
major part of galaxies in this region ($\gtrsim$80 per cent) are found via the 21-cm
\HI\ radio-line (blind) surveys (HIPASS, ALFALFA) or the dedicated \HI\
surveys to probe the content of ZOA \citep{Staveley2016,McIntyre2015}.
Also, the part of ZOA galaxies were identified via the redshift measurements
of extended NIR sources (2MASX) from the Two-micron survey \citep{2MASS}.
Therefore, the subsample of void galaxies in this region is biased to more
or less gas-rich objects.

It is naturally that the completeness of void galaxies in ZOA is much reduced
with respect of that in the other sky regions. Only for a part of them,
the optical counterparts are found. The main numbers for void galaxies in
the ZOA are as follows.
Of the total 1354 void objects, 160 fall in ZOA. Of these 160 galaxies,
69 objects have no optical photometry. The majority of them are either
invisible in the common optical surveys, or are bearly detectable.
The absolute magnitude distribution for the remaining 91 galaxies is
clearly shifted to the larger luminosities. Only 5 galaxies ($\sim$5.5 per cent)
of these 91 have $M_{\rm B} \gtrsim -14.2$ mag, with the median of
$M_{\rm B,median,ZOA} = -16.84$ mag, in comparison to the median of the whole
void 1246 galaxies with known $M_{\rm B}$, $M_{\rm B,median} = -15.22$ mag.
Of them, 373 ($\sim$28 per cent) are fainter than
$M_{\rm B} = -14.2$ mag, conditionally selected as a border between the least
luminous dwarfs.

\subsection{Notes on Nearby Void galaxy clustering}
\label{ssec:clustering}

As far as we aware, the issue of void galaxy clustering was little addressed
earlier. There was a paragraph in our description of galaxy sample in the
nearby Lynx-Cancer void \citep{PaperI}, where it was noticed that around
18 per cent of them form pairs. A more advanced study of this issue was performed
with \HI\ mapping of about 60 void galaxies from the Void Galaxy Survey (VGS)
\citep{Kreckel12}. With the newly \HI\ detected mostly less massive companions
of the main sample void galaxies, the authors found that $\sim$40 per cent of
VGS galaxies (including the new detections) are members of pairs or triplets.

In the current sample, thanks to a much larger statistics, we can give an
independent and more general estimate of clustering frequency of void
galaxies. We marked in the last columns of the Appendix Tables A1 and A2
by '++' all galaxies which either certainly or with the large probability
to be in pairs, triplets, quartets or in a void group NGC428. For the 'inner'
subsample (Table A1) of 1088 galaxies, 190 ($\sim$17.5 per cent) objects form small
aggregates.
For 266 objects from the 'outer' galaxy subsample, presented in Table A2, the
situation is similar. 46 ($\sim$17.2 per cent) of them form pairs or triplets.
Overall, 236 of 1354 Nearby Void galaxies ($\sim$17.4 per cent) appear in the bound
aggregates.

It is interesting to compare this number with that for the general galaxy
sample in the nearby Universe derived in
\citet{LSCPairs,LSCTriplets,LSCGroups}. According to those studies, the
fraction of galaxies within the bound aggregates in the Local Supercluster
comprises of 54 per cent. Thus, the coarse comparison of void galaxy sample
and the whole galaxy sample in nearby Universe reveals the drastic difference
in their local environment.
To estimate more accurate differences in the local clustering of galaxies from
both samples, one needs to form galaxy samples with similar selection
criteria. This will be a task for a more
careful study of the discussed void galaxy sample.

As for the comparison with the clustering results for void galaxies from VGS,
one needs in a sensitive \HI\ mapping of representative subsamples from
the current Nearby Void Galaxy sample. This should bring many new low-mass
void objects, probe void galaxy mass function at the lowest limits, will
probe the galaxy aggregates elongation and probable connection with void
filaments and substructures. The whole sky sensitive \HI\ mapping with the
next generation radio astronomy facilities - SKA and ngVLA, and their
pathfinders Apertif, ASCAP and MeerKAT will greatly advance our understanding
of the low-mass galaxy population in the nearby Universe.

\subsection{Preliminary summary of \HI\ data}
\label{ssec:HI}

For 980 galaxies from the Nearby Voids, we found data on \HI\ fluxes in LEDA
and in the literature.
Most of them come from two blind \HI\ surveys: HIPASS and ALFALFA.
One of the important related questions is how much gas the void galaxies
have in comparison to similar galaxies in denser environments.
There are direct observational indications on the enhanced gas mass fraction
in void galaxies. But, on the one hand, they are not numerous due to small
statistics for nearby voids \citep{Paper6}. On the other hand,
for more luminous galaxies in more distant voids (e.g., VGS, \citet{Kreckel11,
Kreckel12}), the situation with the gas mass fraction with respect of the
control ALFALFA sample is less certain despite a substantial number of found
the lower  luminosity gas-rich objects.

The recently published final ALFALFA release \citep{ALFALFA18} of about 31500
\HI-bearing
galaxies should significantly advance our understanding of this issue in both
the lowest mass void galaxies and the more distant and more massive void
objects.

\begin{figure}
 \centering
\includegraphics[angle=-90,width=7.5cm,clip=]{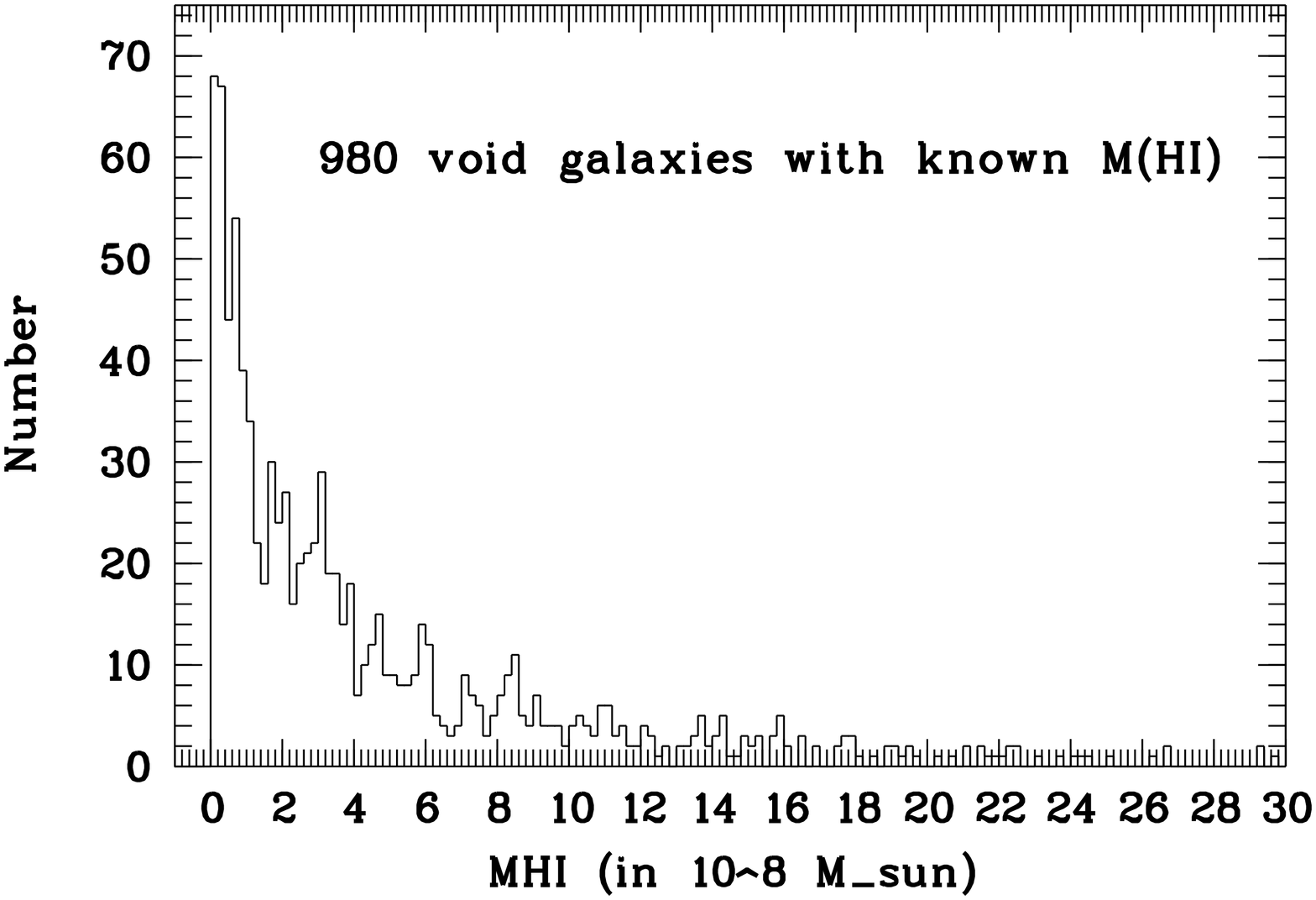}
\includegraphics[angle=-90,width=7.5cm,clip=]{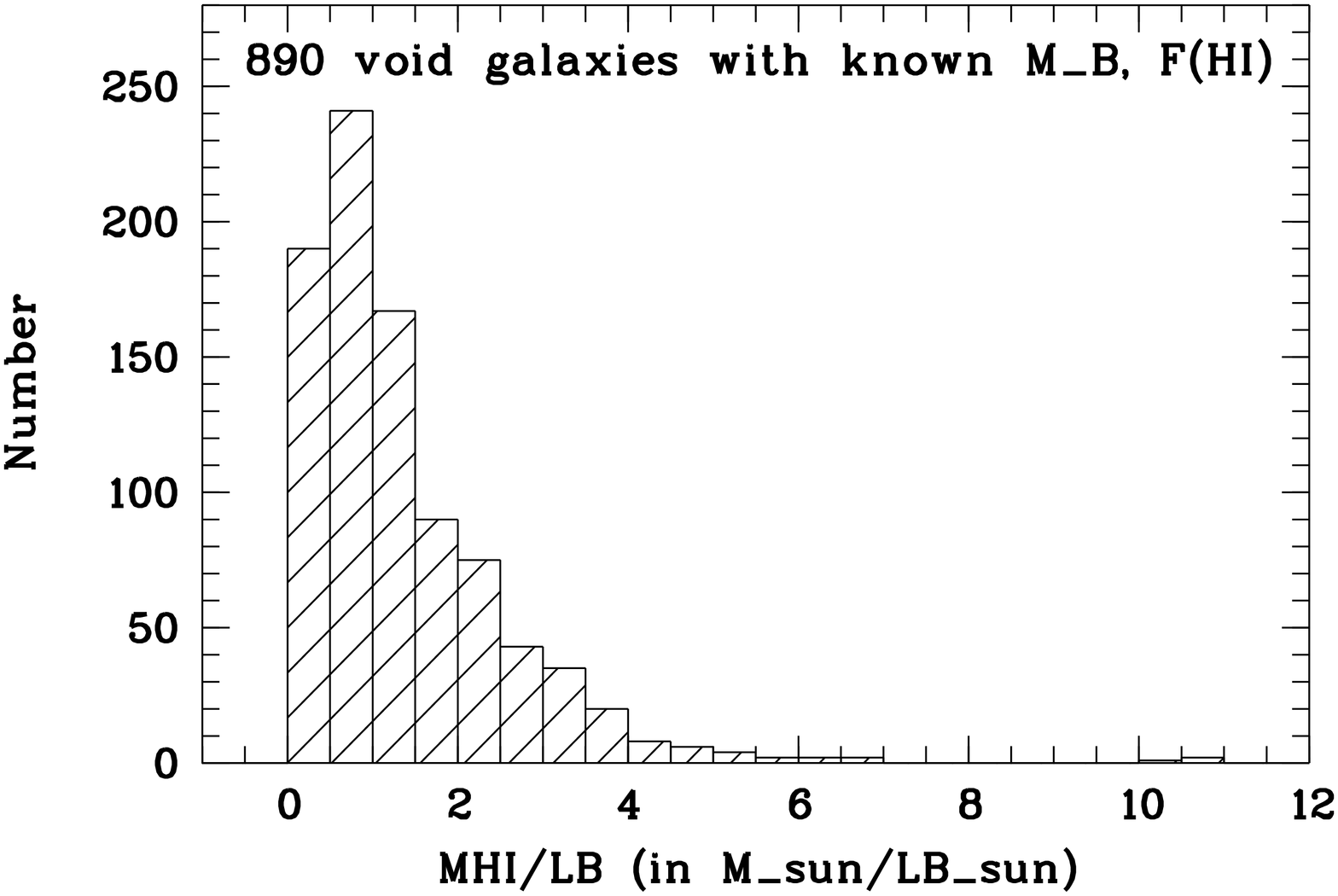}
\includegraphics[angle=-90,width=7.5cm,clip=]{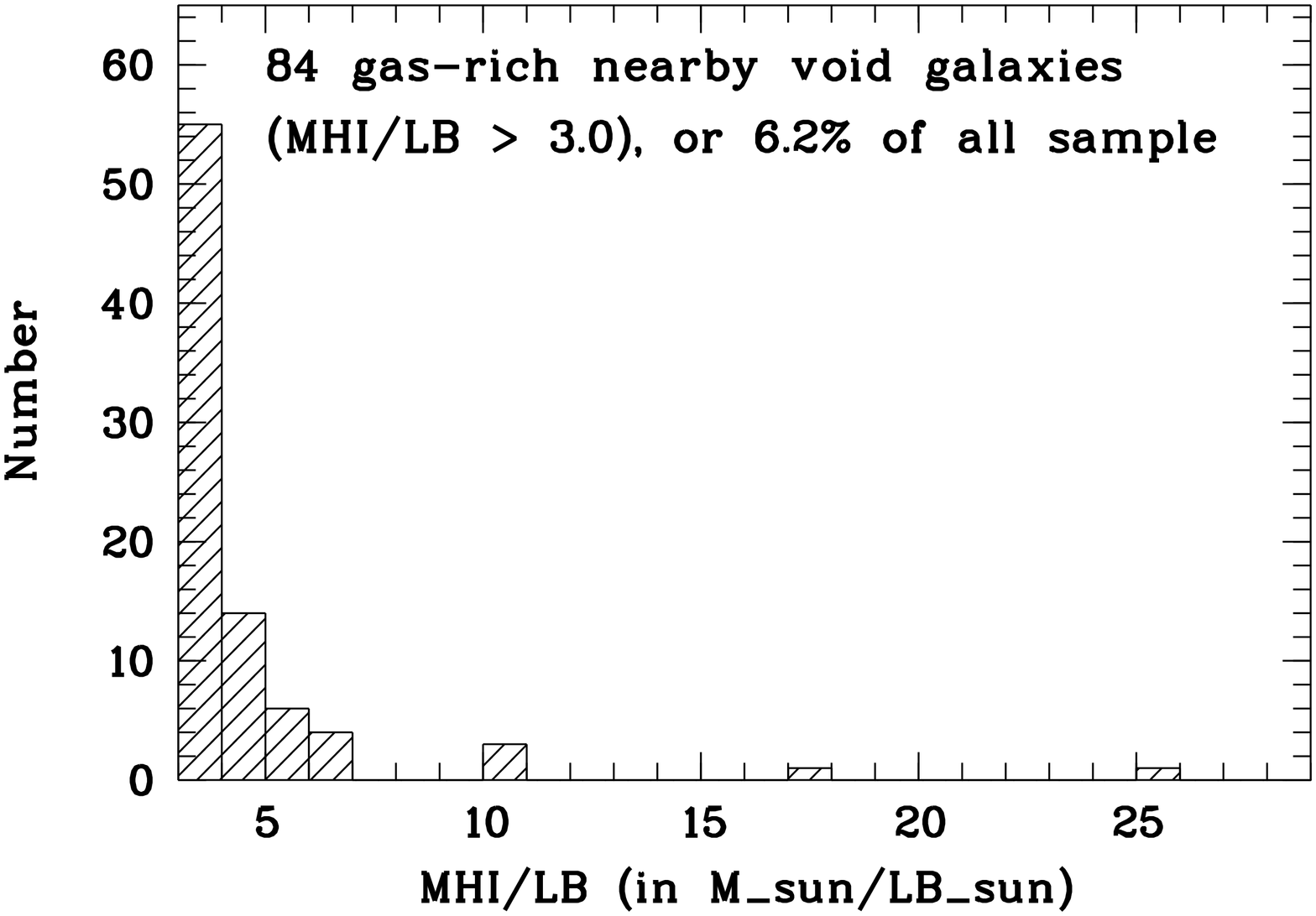}
  \caption{\label{fig:hist_HI}
{\bf Top panel:} Distributions of all 980 nearby void galaxies with \HI\
  detections on mass of \HI\ (in units of 10$^8$~M\sunn). 19 galaxies with the
  largest M(\HI), of 30--52, fall out of the plot.
{\bf Middle panel:} Distribution of all 890 nearby void galaxies with known
$M_{\rm B}$ and M(\HI) on the ratio M(\HI)/$L_{\rm B}$ (in solar units).
{\bf Bottom panel:}  similar distibution for the most gas-rich 84 void galaxies
with M(\HI)/$L_{\rm B} > $3.0 }
\end{figure}

In Fig.~\ref{fig:hist_HI} we present the distribution of all 980 nearby void
galaxies with the available \HI-flux detections. In the top panel the
histogram of the sample M(\HI) is shown, in units of 10$^8$~M\sunn. It is
very broad, from 0.003 to 52, with the median of 2.85. 34 galaxies have
M(\HI)/($10^8$ M\sunn) $<$ 0.1, and
35 more have M(\HI) in the range (0.1--0.2)~10$^8$~M\sunn.
In the middle panel we show the distribution of parameter M(\HI)/$L_{\rm B}$
(in solar units) for all 890 void galaxies with known $M_{\rm B}$ and M(\HI).
It also is very broad, from 0.03 to $\sim$26, with the median of 1.03.
In the bottom panel we show the closed-up part of this histogram for 84 the
most gas-rich void galaxies. One can notice that this distribution steeply
falls for M(\HI)/$L_{\rm B} >$ 4--5. Only a handful of void galaxies are found
with this parameter to be more than 5. Most of them were already found in
the recent publications. They include, in particular, J0723+3622 and
J0723+3624 with M(\HI)/$L_{\rm B}$ = 10.7 and 25.6 respectively \citep{CP13},
J0706+3020 with M(\HI)/$L_{\rm B}$ = 17.1 \citep{UGC3672A},
AGC198691 (M(\HI)/$L_{\rm B} \sim$7 \citep{Hirschauer16}, J0110-0000
(M(\HI)/$L_{\rm B} \sim$6.5), J2104-0035 (M(\HI)/$L_{\rm B} \sim$4.4
\citep{Ekta08}.

We should comment at this place on caution to use the LEDA parameters to
separate the most gas-rich galaxies without the careful additional checks
of the adopted database values. In particular, in the first pass to build
this distribution, we found substantially larger number of very gas-rich
objects than it is shown here. To check their reality, we examined the
subsample of 56 void galaxies with the highest M(\HI)/$L_{\rm B} >$ 3.
15 of them, mainly the most gas-rich, appeared faulse due to unproperly
adopted $B$ magnitudes in LEDA database. In particular, when there are
multiple measurements, the adopted magnitudes are often the old ones, with
the low accuracy and typically are underestimated. In cases of the only $B$
magnitude estimates, obtained from photoplates, there exist new data in the
literature with CCD photometry which
differ substantially in the direction of brighter magnitudes. Correction can
be as high as $\sim$1 -- 2.5 mag, or a factor of 2.5--10 in reduction of
the automatically derived parameter M(\HI)/$L_{\rm B}$.
We also checked for two dozen of this group void galaxies, the model
magnitudes presented in the SDSS database, performing own aperture photometry.
For the great majority of the examined galaxies, the total magnitudes from
aperture photometry appeared systematically brighter, with the typical
difference of $\sim$0.5 mag.

We have checked only the most obvious cases which resulted in the artificially
elevated M(\HI)/$L_{\rm B}$. This allowed to fix the largest deviations in
LEDA database. It is not clear, how often the smaller underestimates of
$M_{\rm B}$ occur which can somewhat affect more numerous galaxies
with more typical values of M(\HI)/$L_{\rm B}$. This can take a special
massive study. However, for goals of statistical analysis of various samples,
one should keep in mind that real $M_{\rm B}$ in LEDA for samples of nearby
galaxies can be partly biased to the fainter luminosities.

\begin{figure*}
 \centering
\includegraphics[angle=0,width=14cm,clip=]{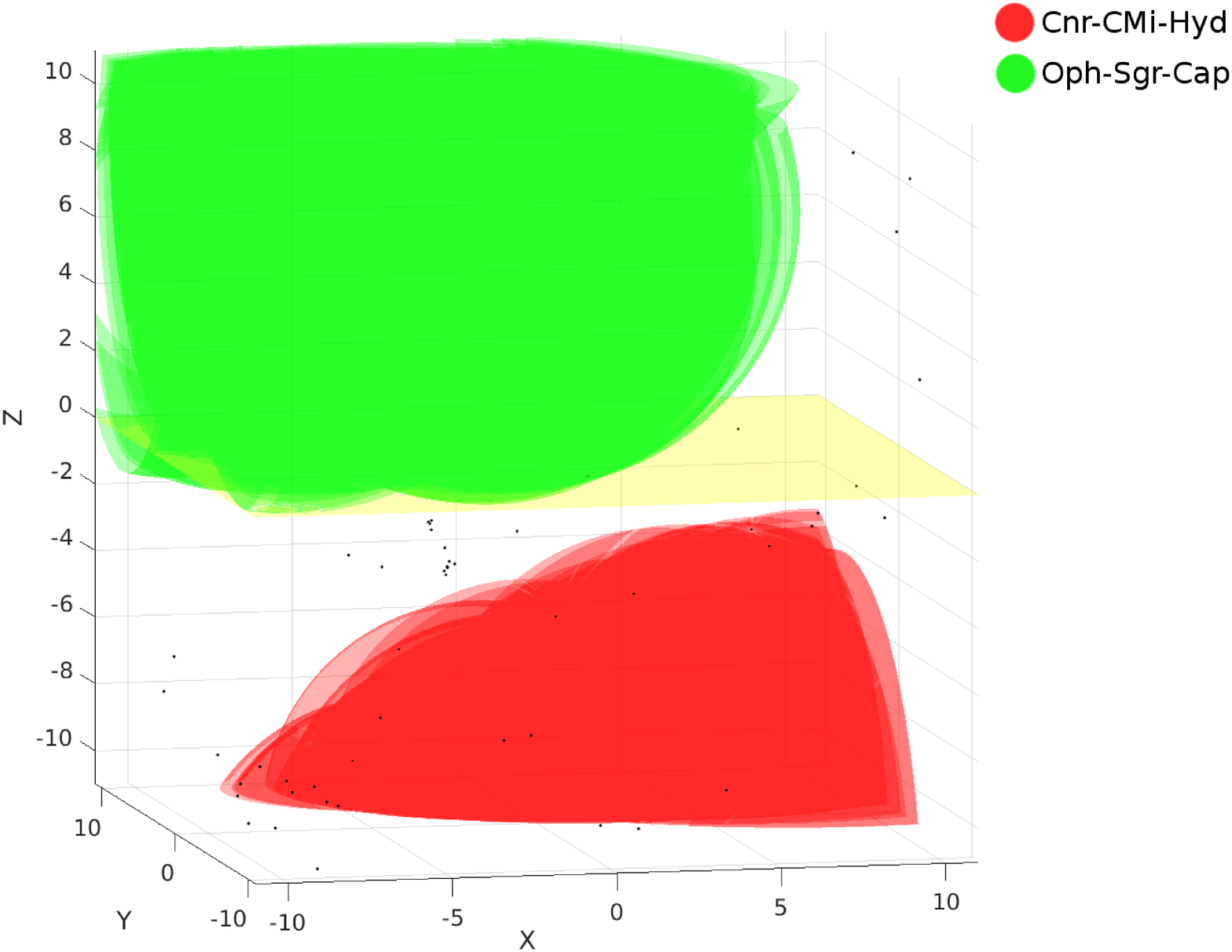}
\includegraphics[angle=0,width=14cm,clip=]{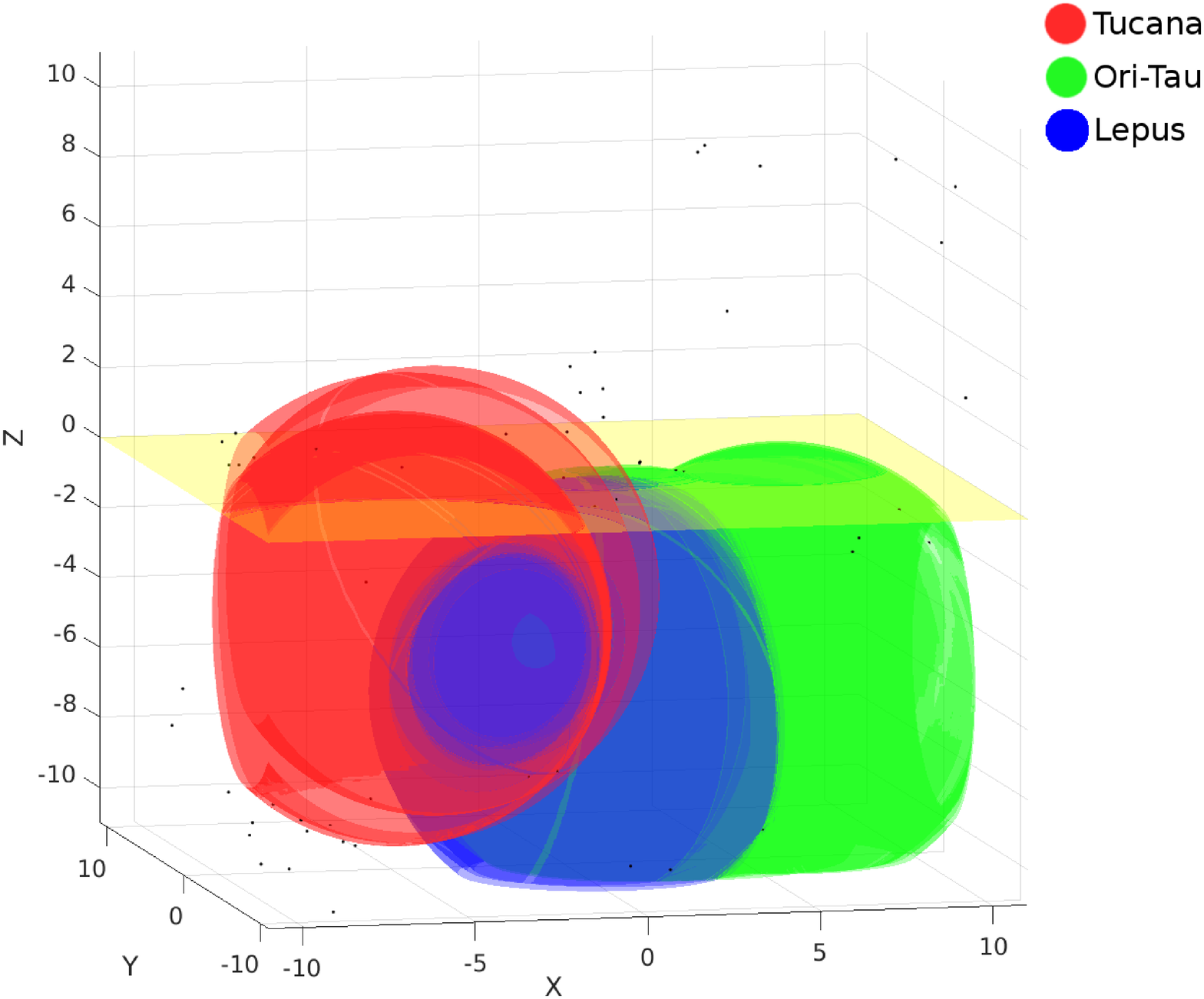}
  \caption{\label{fig:voids_close.3D} 3D view in supergalactic coordinates
of several the nearest voids which occupy the large fraction of the Local
Volume. {\bf Top panel:}  Close parts of the large voids
Ophiuhus--Sagittarius--Capricornus (green) at positive SGZ and
Cancer--Canis-Minor--Hydra (red) at negative SGZ. The position
of plane $SGZ=0$ is shown in aid to eye.
{\bf Bottom panel.} Similar view of three smaller close voids:
Tucan (green), Ori-Tau (red) and Lepus (blue).}
\end{figure*}

\subsection{Nearest voids within the Local Volume}
\label{ssec:local_volume}

The issue of the nearest void galaxies, and in particular of those belonging
to the Local Volume, is worth to briefly discuss since their proximity gives
them an additional advantage to be studied in a more detail.
As can be seen in Fig.~\ref{fig:voids_close.3D}, a part of the nearest voids
falls into the sphere with $R = 11$~Mpc. This is the definition of
the updated version of the Local Volume related to the sample of the
Updated Nearby Galaxy Catalog (UNGC) \citep{UNGC13}.
The most updated version of this catalog includes
1153 galaxies.\footnote{https://www.sao.ru/lv/lvgdb}

\begin{figure}
 \centering
\includegraphics[angle=-90,width=7.5cm,clip=]{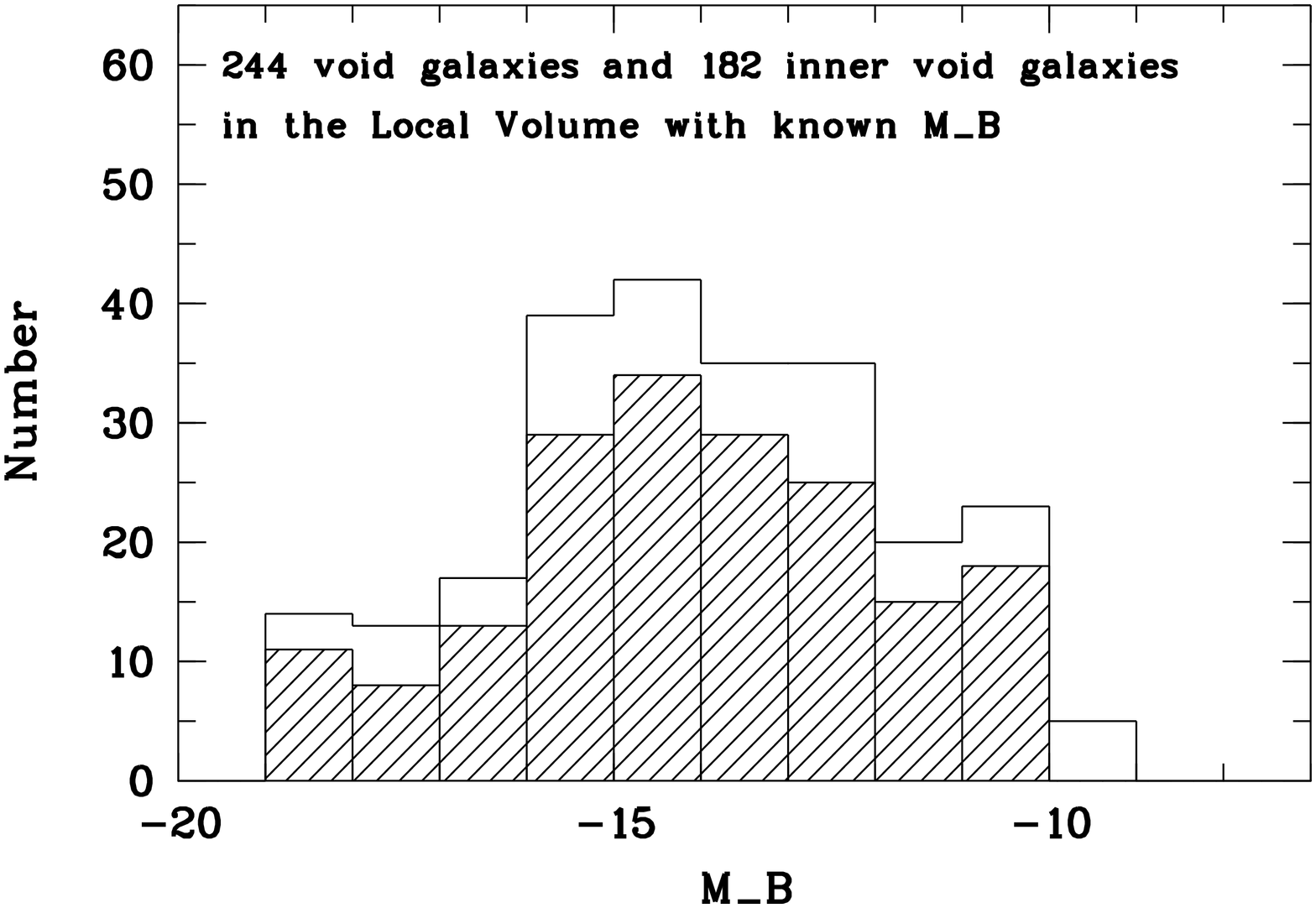}
\includegraphics[angle=-90,width=7.5cm,clip=]{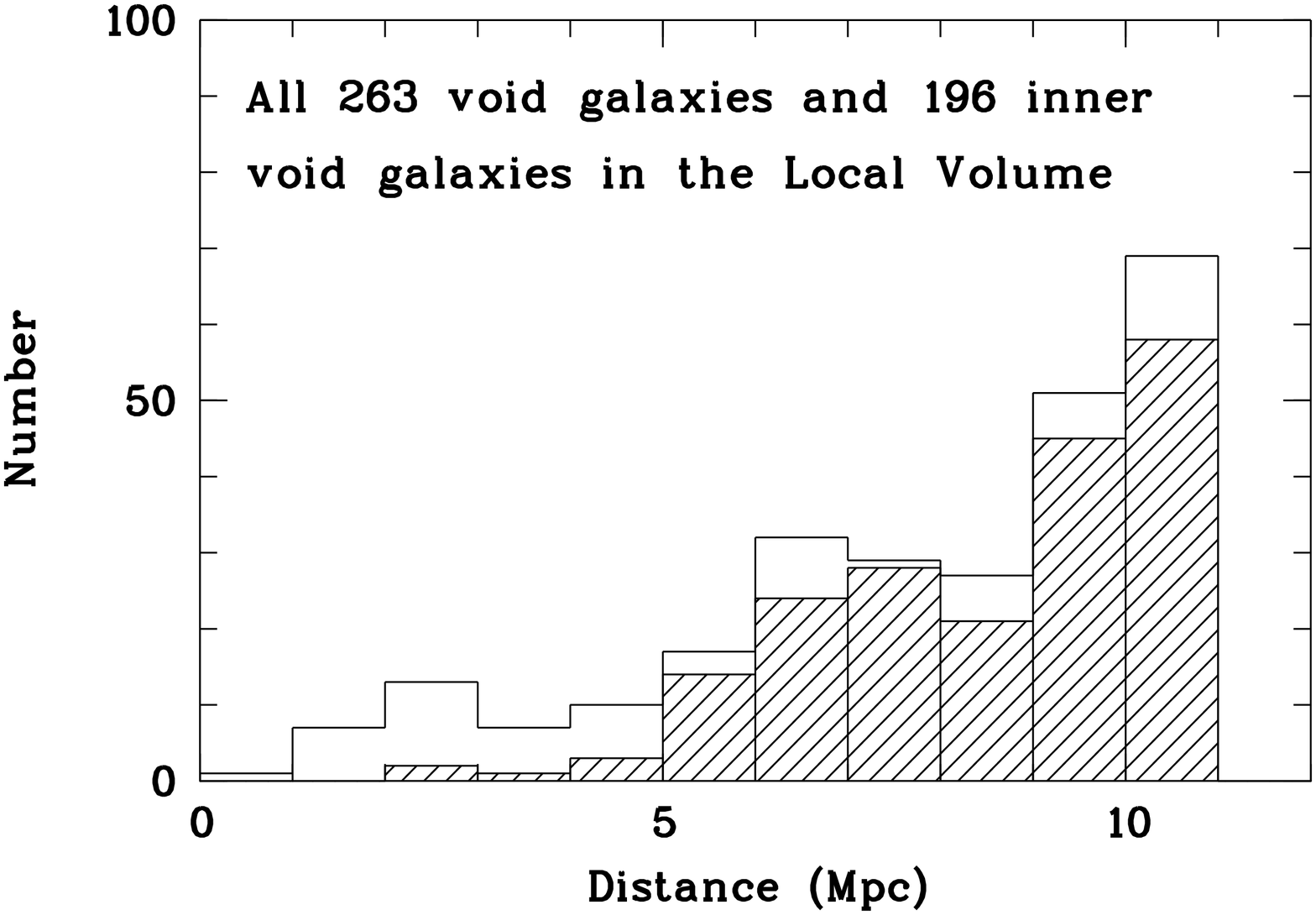}
\includegraphics[angle=-90,width=7.5cm,clip=]{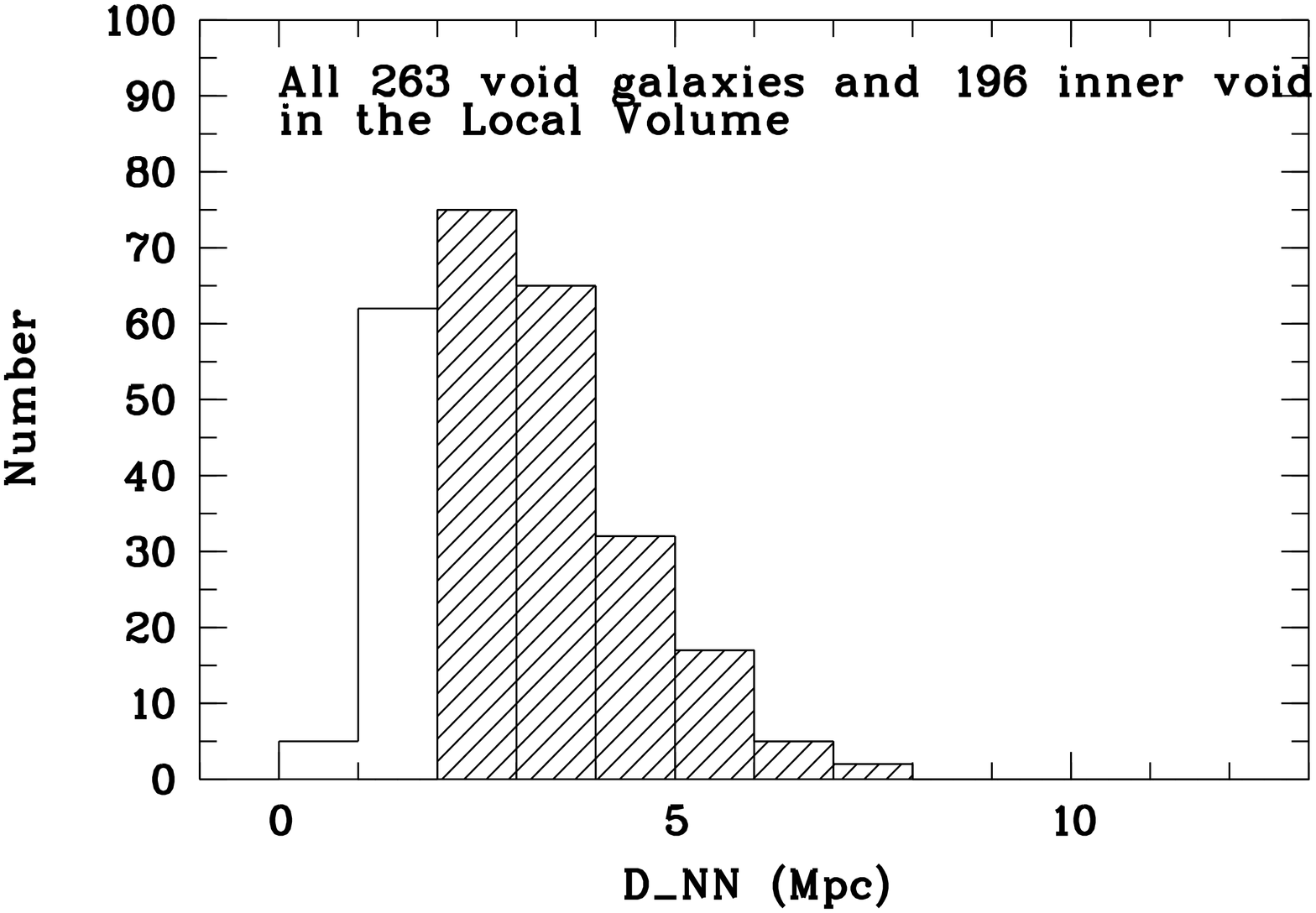}
  \caption{\label{fig:histo_LV}
{\bf Top panel:} Distribution of all 243 and 181 'inner' (shown by hashing)
void galaxies, residing within the Local Volume ($D <$ 11~Mpc), on $M_{\rm B}$
(with known $B_{\rm tot}$),
{\bf Middle panel:}  similar distibution on the distance (in Mpc) for all 263
and 195 'inner' (hashed) void galaxies in the Local Volume,
{\bf Bottom panel:} similar distribution on $D_{NN}$, the distance to the
nearest luminous object, for all 263 and 195 'inner' void galaxies in the
Local Volume. }
\end{figure}

The distributions of void galaxies residing in the Local Volume on
$M_{\rm B}$, distances and $D_{NN}$ are shown in Fig.~\ref{fig:histo_LV} (hashed
part marks 'inner' void subsample).
The number of void galaxies falling within the Local Volume (LV) is currently
263. That is, the nearby void galaxy population comprises $\sim$23 per cent
of all known galaxies in the LV. Of them, 195 belong to the 'inner' void
galaxy population, while the remaining 'outer' 68 galaxies
are situated closer to the void boundaries. The fraction of the LV
void galaxies in the regions with the large Milky Way extinction
(adopted here as $A_{\rm B} > $ 1.0 mag) is a substantial:
 $\sim$12 per cent (29 objects, of them 23 galaxies of the 'inner' subsample
and 6 - of the 'outer' subsample).
In summary, the sample of the inner void galaxies counts $\sim$160
unobscured objects,  or $\sim$14 per cent of the whole number in this volume.
This gives us the good prospects of their deep
studies, including the possible peculiarities of their evolution via the
resolved stellar populations with the {\it HST} and the future space
telescope {\it JWST}.

In particular, it is interesting to compare the number of void galaxies
within the Local Void region (in our scheme, this is No.~21, Oph-Sgt-Cap).
It approaches the Local Sheet and comes close to the Local Group. Sixteen
Local Void galaxies were identified recently by \citet{LocalVoid11}.
Our subsample includes 89 'inner' galaxies in comparison to the above
16 galaxies in their work. Of them, 82 galaxies are new Local Void objects.
Of these 82, 26 galaxies have $D < 11$~Mpc, that is reside in the Local
Volume.

Additional 32 galaxies from the 'outer' subsample also
fall  within the Local Void. Of them, 28 are new Local Void galaxies.
17 of these galaxies have $D <$11~Mpc.
A half of these 32 objects have $D_{\rm NN} > 1.6$ Mpc, and thus, they also
can be studied as residing rather deep in the Local Void.
In total, 36 galaxies in the Local Void fall in ZOA
($A_{\rm B} > 1.0$ mag). However, only 14 of them fall in regions with
$A_{\rm B} > 2.0$ mag.

One of the important advantages in the study of the nearest voids is the
opportunity to measure the peculiar velocities of void galaxies due to void
expansion via taking their independent good accuracy distance estimates with,
e.g. TRGB method based on the HST images. The potential of this direction was
recently successfully
demonstrated by \citet{Rizzi2017} for two dwarfs within the Local Void at
distances of $\sim$8.4 and $\sim$7~Mpc, with the estimated galaxy velocities
away from the void center to be $\sim$270 and $\sim$370~\kms. The advancement
in the nearest future of TRGB distance measurements with the new JWST 6-m
space telescope and in the close perspective with such ground-based giants
as E-ELT and TMT will allow to extend such study to the outer border of
the Nearby Void Galaxy sample.

\subsection{Some unusual galaxies residing in the nearby voids}
\label{ssec:unusual}

As noticed in Introduction, the deep unbiased study of faint LSBDs in the
Lynx-Cancer void resulted in the discovery of about ten
blue very gas-rich and extremely metal-poor dwarfs. They represent
$\lesssim$10 per cent of the total void galaxy sample. However,
their fraction increases till $\sim$30 per cent for the lowest
luminosity void LSB dwarfs. Their observational properties are consistent
with the hypothesis that such unusual void dwarfs represent the youngest
local galaxies. In particular, they can be among the local very young
galaxies (VYGs) defined by \citet{Tweed2018} as galaxies which formed more
than a half of their stellar mass during the last 1~Gyr. In particular,
they may be more quiscent analogs of extreme metal-poor star-forming
galaxies such as J0811+4730 with $Z = Z$\sunn/50 \citep{Izotov18}.

One of the tasks of formation of the Nearby Void Galaxy sample was its
potential to search for new unusual least evolved objects and to increase the
sample of such galaxies several fold. This will allow us to study their
properties statistically and to address their origin and nature. At this
stage, having for the cataloged void galaxies only the general parameters
from the public databases, we can only make a preliminary selection of
candidates to such unusual galaxies for the subsequent checks of their
crucial properties. The subsample of about three dozen such candidates
selected for the additional observations will be presented in the accompaning
paper. Just as examples, we mention a couple of such new objects.

The first one, J0110-0000 (AGC411446) is a very blue faint
($M_{\rm B} = -11.3$ mag) and gas-rich [$M$(\HI)/$L_{\rm B} \sim 6$] dwarf
at $D \sim$ 17~Mpc, with the preliminary estimate of 12+$\log$(O/H) = 7.01
dex. Its stellar mass, estimated with the commonly used light-to-mass ratio
from \citet{Zibetti09}, comprises less than 1 per cent of the whole baryon mass
(Pustilnik et al., 2018, in preparation). From the sum of its properties,
this object is a good representative of the similar objects found in the
Lynx-Cancer void sample, a candidate VYG and the record low-metallicity
galaxy in the nearby Universe. Its O/H is very close to that of the similar
void  dwarf AGC198691 (Leoncino) \citep{Hirschauer16}.

The second galaxy, J1349+3544 (AGC239144) at $D \sim$ 20~Mpc, is also very
blue, gas-rich [$M$(\HI)/$L_{\rm B} \sim 3.2$] and faint
($M_{\rm B} \sim -12.5$). From the intermediate quality spectrum of its
faint \HII\ region, obtained recently with the SAO 6-m telescope, we estimate
its tentative O/H at the level of (O/H)\sunn/50. When this preliminary value
of O/H is confirmed, this galaxy will appear one more candidate to void VYGs.

\subsection{SDSS Stripe~82}
\label{ssec:S82}

The SDSS Stripe~82 covers 275 sq.deg within the RA range of 310\degr\ to
59 \degr\  (or $20^h40^m - 04^h56^m$) and Dec = --1.25\degr\ to +1.25\degr.
Its apparent magnitude limit is deeper by $\approx 2^m$ than the standard
SDSS limit \citep{Stripe82}. Of our total void galaxy list, 22 objects fall
inside the region of the Stripe~82. This presents us an opportunity to study
nearby void galaxies to a much fainter surface brightness level. Besides,
this database allows us to search for fainter companions near already known
void dwarfs. This is a way to extend the census of void galaxies to the
lowest mass objects. This approach we already tested successfully on the
Lynx-Cancer void sample.

In particular, in this region we identify near the center of Void No.3
the unusual group with the dominant spiral NGC428 with $M_{\rm B} = -19.2^m$.
It includes seven fainter galaxies with $M_{\rm B}$ in the range of --11.3$^m$
to --15$^m$. At least three of them are galaxies with very low O/H. These
are UGC~772 with 12+$\log$(O/H) = 7.15--7.32 \citep{izotov2012}, IC~88 with
12+$\log$(O/H)=7.56 \citep{Kniazev18} and the mentioned above J0110-0000
with 12+$\log$(O/H)$\sim$7.0. The group represents a clear density peak
within a void, and according to the models of void substructure by
 \citet{Sheth2004, Goldberg2004, AragonCalvoSzalay2013, Rieder13, Rieder16},
it may be a node of several void filaments.
Thus, its very deep mapping in \HI, similar to suggested in the project
MHONGOOSE \citep{MHONGOOSE}, may be a mean to probe a tenuous gas in the
related filaments.

\subsection{Void small-scale structures}
\label{ssec:void_structure}

One of the important issues of the modern cosmology paradigm is the nature
of the ubiquitous dark matter component. As a complementary to the widely
adopted models with only Cold Dark Matter (CDM), several groups elaborate
the effect of various kinds of Warm Dark Matter (WDM) mixed at different
proportions with the CDM component. The main difference of WDM models with
the standard CDM models appears at 'small' scales, on which the streaming of
WDM particles should wash out structures \citep{Tikhonov09,Angulo2013}.
To date, the firm predictions of void structure at small scales for various
fractions of WDM are absent. However,  the progress in  the N-body
simulations with the increasing  mass resolution allows us to hope
to nearby future advancement  in the theoretical predictions of void
properties at the smallest scales comparable to $\sim$1--2 Mpc,
e.g. \citet{AragonCalvoSzalay2013}.

To date the issue of small scale filaments in voids based on observations
is studied very little. Only a couple of examples mentioned in
Sect.~\ref{ssec:nearby_voids} was addressed.
To be able to compare different predictions of DM scenarios including
various fractions of WDM (e.g. \citet{Yang2015}),
observers need to deal with the void small-scale
structures, which in turn can be probed only with a sufficiently large
number density of lower mass halos (galaxies).  This situation dictates
the need to pay attention to the deeper probes of the void environment in
the nearby Universe, and in particular to the space of the Local Supecluster
adjacent the Local Volume. With the presented Nearby Void Galaxy sample,
we pursue the first steps in creating this cosmologically important benchmark
galaxy collection. We expect the further substantial increase of this sample
thanks to results of the future wide area deep \HI\ surveys with the SKA and
its pathfinders Apertif, ASKAP and MeerKAT as well as the New Generation VLA.

\subsection{Limitations of the void galaxy catalog}
\label{ssec:limitations}

As described in Sec.~\ref{ssec:procedure}, in the process of selection of
empty spheres, we adopted the lower limit for their radii of 6.0 Mpc. This
decision was motivated, on the one hand by the goal to limit the voids'
sizes by more or less commonly used sizes. On the other hand, this allows one
to prevent the effect of 'percolation' which joins many larger voids into
connected regions via the tunnels of smaller voids. This arbitrary limit is
somewhat artificial, reflecting the authors' intention to describe the nearby
lowest density regions as a collection of more or less individual voids.

The smaller voids, and in particular, some smaller 'pockets', adjacent the
larger ones, occupy the significant fraction of the volume not included into
presented void sample. Their void dwarf population certainly can be similar
to the population of larger voids. We mention two known galaxies falling into
the volume under consideration which are very gas-rich, low-metallicity
objects, and, hence, are similar to many void galaxies. The first one
is And~IV \citep{Kar2016,AndIV} at $D \sim$7~Mpc falling in a minivoid with
the total size of $\sim$7~Mpc. The second example is a very gas-rich pair of
a metal-poor LSBD and an \HI-cloud without visible optical counterpart,
HI~1225+01 \citep{Salzer91}. This unusual pair resides at the distant
periphery of the Virgo Cluster, at 3--6 Mpc \citep{Salzer92}, depending on its
real distance (10--20 Mpc), which is poorly determined due to the object falls
in the 'triple-valued region' of the Virgo Cluster velocity field. However,
its more 'local' environment on the scale of $\sim$1--2~Mpc is devoid of
luminous galaxies \citep{Salzer92} and forms a kind of a low-density pocket.

The real void-intervoid web structure of matter distribution
is of course much more complex than we approximate it in this paper.
The future work aimed in a more precise description of the structure of the
nearby region of the Universe will extend our approximate description and
will allow us to follow better finer details and their possible effects
on galaxies within these low-density regions.

It is worth to mention on the substantial inhomogeniety of the current nearby
void galaxy sample on the depth along the celectial sphere. This is related to
both the North-South asymmetry of the redshift data and the substantial loss
of depth in voids falling to the sky regions at low galactic latitidues.
The North-South asymmetry is caused by the substantial contribution to
the northern hemisphere of void galaxy sample from both the SDSS redshifts
and ALFALFA redshifts which allow us to identify the fainter galaxies
with respect of those in the Southern hemisphere.

\subsection{Comparison with similar works}
\label{ssec:similar.work}

It is worth to compare our 'Nearby Void Galaxy' sample with several similar
published samples with the emphasis on the main differences of our approach.
As already was mentioned above, the main idea was to extend to a much
larger volume the galaxy sample formed by us earlier in the
nearby void in Lynx-Cancer \citep{PaperI}. That sample has the very
important advantage over several other known samples thanks to its
significantly deeper probe of the faintest and the least massive void
galaxy population. The similar work was recently performed for a somewhat
more distant void by \citet{Kniazev18}. The study of galaxies in both
voids resulted in the similar conclusion on the slower void galaxy evolution.

Other studies of more or less similar samples which were published recently,
include papers by \citet{Elyiv2013} (voids within 50 Mpc and their dwarfs) and
\citet{Bradford15} (isolated dwarf galaxy sample based on the SDSS database).
The papers by \citet{Moorman15,Moorman16} deal with the ALFALFA-SDSS galaxies
in voids
determined by \citet{Pan2012} on the SDSS DR7 galaxy sample. Since their
nearest voids are situated at $D \sim$ 50~Mpc, the study of \citet{Moorman15}
is limited by the more luminous dwarfs. The Void Galaxy Survey (VGS) by
\citet{vandeWey09},
\citet{Kreckel12} is devoted to the detailed study of 60 galaxies situated
in the centers of large voids of the SDSS footprint. Similary, it deals
mostly with more distant voids (average distance of $\sim$80~Mpc) and
therefore only a few their sample galaxies have $M_{\rm r} > -16.0$.

\citet{Elyiv2013} find only about 70 dwarf galaxies residing in their voids
with  $D$ up to $\sim$50~Mpc. About 1/3 of them fall within $D <$ 25~Mpc.
Most of those 24 galaxies reside in our voids.
They contribute only a few per cent to the number of our void galaxies with
$M_{\rm B} > -16.0$ what roughly corresponds to their $M_{\rm K} > -18.4$.

One of the closest samples of isolated galaxies is the so called Catalog of
Local Orphan Galaxies (LOG, $\sim$500 galaxies with $z < 0.01$) by
\citet{LOG2011}.
Due to their local isolation criteria, the number of common objects of LOG
and nearby void sample is relatively small. It can be examined by the
statistical study of evolutionary parameters for both samples, which of
factors: local or global isolation, are more important. As an example
indicating the probable prevalence of the global low density (aka void
type population), we refer to several the least evolved void LSB dwarfs
mentioned in Introduction, which enter to physical pairs and triplets. This
membership and the related tidal interactions
can be an appearance of the recent (time-scale of $t \lesssim$1--2~Gyr)
an aggregate assembly. This, in turn, acts as a trigger of a relatively short
time-scale recent episode of star formation in void protogalaxies
with the globally delayed evolution.

\citet{Bradford15} also deal with the sample of isolated galaxies
selected from the SDSS database but in a much larger volume with $D$ up to
250~Mpc. Their criterion of isolation of $D_{\rm NN} >$ 1.5~Mpc from
luminous galaxies with $M_{\rm K} < -23.0$ is weaker than our
($D_{\rm NN} \geq $ 2.0~Mpc for galaxies with $M_{\rm K} < -22.0$) and
ignores the large-scale environment. Of their sample of 1715 isolated
galaxies, only $\sim$50 fall into the volume with $R <$25.0~Mpc, which is
the subject of our study. It will be interesting, after accumulation of
the sufficient data for our Nearby Void Galaxy sample (including \HI-data),
to compare its properties with those of \citet{Bradford15} sample. This
probably will help to disentangle the effect of low-density environment and
that of the local isolation.

\subsection{Issue of wrong objects in the initial catalog}
\label{ssec:wrong}

As described in Sec.~\ref{sec:voidgal}, from the sample of 2075 initially
selected objects, we removed 361 satellites of luminous galaxies or members
of the related groups as well as probable representatives of several large
galaxy concentrations. Of the remaining 1714 selected in LEDA objects, which
fall in the interior of the found empty spheres, there was rather large
amount of entries with either unknown photometry (in total 356),
or with the adopted very faint magnitudes which resulted in the absolute
$B$-magnitudes of $-5$ to $-11$ for galaxies outside the Local Group or
nearby groups (in total 177).
We performed the careful checks of these 533 as well as other doubtful
objects to find possible errors in the original catalog data.
Indeed, we identified 312 entries with various kind mistakes
which we excluded from the final void galaxy sample presented in Tables A1
and A2 of Appendix.

The excluded 312 objects fall in the next categories, with the respective
numbers in parentheses.
a). Galactic stars with the radial velocities of several hundred up to about
a thousand \kms\ (91). A part or all of them can be non-recognized runaway and
hypervelocity stars which are expected to be rather numerous in the Milky Way
\citep[e.g,][]{Brown15,Marchetti18}.
b). Stars with the same range of radial velocities, projected near the central
parts of background distant galaxies (40).
c). Galaxies with the SDSS spectra which in the latest versions of SDSS
database have red-shifts much larger than the adopted in LEDA, or marked
in LEDA as uncertain ones (81).
d). Small parts of galaxies (usually \HII-regions), identified as independent
extragalactic objects (42). e). Quasars (7). f). Various artifacts, like
strays from bright stars (10), g) PNe and GC (4), h) doubtful \HI\ sources
(6).
Around 30 more of the original LEDA entries were excluded on various
additional reasons.
One of particular cases, the object with the unusual morphlogy, PGC3097159
(SDSS
J061431.3+073424), appeared in HyperLEDA with $V_{\rm hel} = 971\pm24$~\kms.
According to our BTA 6m telescope spectrum, this is the Milky Way planetary
nebula with the real $V_{\rm hel} = 27\pm24$~\kms\ as derived from the
observed wavelengths of H$\alpha$ and the strong line [NII]$\lambda$6583.

Besides, when the preliminary version of the Nearby Void galaxy sample was
formed, we realized that for part of them the cited errors of $V_{\rm hel}$
are too large, what hampers the real placement of an object within voids.
We excluded additionally from the void sample 48 galaxies with the cited
$\sigma_V \geq$ 150~\kms.

These findings indicate that in catalogs created mostly via automated
procedures, there exists a sizeable fraction of garbage. As for the formation
of clean samples of galaxies in the nearest regions adjacent the Local
Volume, the contamination with the Milky Way stars and parts of galaxies with
multiple \HII-regions can be the main factors of incorrect identification.

\section{Summary and conclusions}
\label{sec:sum}

In this paper we use the most updated version of HyperLEDA galaxy
collection in the nearby Universe to form the large and representative
sample of galaxies residing in the Nearby Voids, namely, the voids which
we identified within the sphere of $R <$ 25 Mpc. We made several important
improvements with respect of the previous works. They include the following.

\begin{itemize}
\item
The use of the definition of luminous galaxies delineating the voids based on
their absolute $K$-band magnitude. This strongly eliminates the effect of
the Milky Way extiction and, thus, allows us to expand the sample of
delineating luminous objects to the whole sky. This allows us to look for
nearby voids without the artificial break of the search areas by the Zone
of Avoidance.
\item
As it is known from the recent studies, there exists the global motion of
the Local Group (within the Local Sheet), which combines both the attraction
to the center of the Local Supercluster and the repulsion motion of the Local
Sheet as an outer/border part of the giant Local Void,
with the total velocity amplitude of $\sim$350~\kms. We take this motion
into account to correct distances of all galaxies in the volume under the
question, for which we have no independent good accuracy distance estimates.
This allows us, to first approximation to treat all used galaxy distances
(velocity independent and the model-corrected) in a self-consistent manner,
giving the confidence to the formed void galaxy sample.

However, for
the future more accurate description of voids and galaxy distribution in the
studied volume, one needs in more detailed models of peculiar velocities in
the various subregions based on the TRGB-type distances. This step will
require the substantial increase of mass TRGB distance determinations to
the typical distances of 25--30 Mpc with the help of {\it JWST} and the ground-based
giants {\it E-ELT} and {\it TNT}.

\item
The HyperLEDA database for the subset of objects with the radial
velocities of below a thousand \kms\  appeared to contain a substantial
fraction of 'wrong' entries (stars, stars projected close to a distant
galaxy center, galaxies with unreliable redshifts, faint parts of brighter
galaxies, QSOs, various artifacts) which contaminate the void galaxy sample.
We carefully checked all doubtful cases and cleaned 354 such wrong entries
from the intermediate sample of 1708 objects. The latter number left of the
original galaxy sample (2075) cleaned from 319 galaxies, identified as the
(distant) satellites of the luminous galaxies delineating the Nearby Voids,
and 48 objects with low velocity accuracy. We do not
exclude, however, that for the current sample of 1354 nearby void galaxies,
a few entries remain non-recognized as wrong.
\end{itemize}

We describe the procedures to form a sample of 25 nearby voids as a set of
detached complexes of several lumped empty spheres (with the full size of
individual voids of 13--35 Mpc) delineated by 'luminous' objects (galaxies
and typical groups).
Altogether, within these 25 voids we find 1354 galaxies with the
number of galaxies per void from 4 to 129. One of the important
applications of this sample can be tracing of the internal small-scale
structure of voids to test the basic assumptions on the Dark Matter nature.
Of all nearby void galaxies, 1088 reside deeply in the voids that makes
them suitable to study the effect of void environments on galaxy formation and
evolution.

Summarising the results and discussion above, we draw the following
conclusions:

\begin{enumerate}
\item
We present the results of the nearby void identification in the close part
of the Local Supercluster (at $R <$ 25~Mpc). Total 25 voids with the major
size from $\sim$13 up to $\sim$35 Mpc are separated and discussed in relation
to the previous works.
\item
The sample of galaxies residing in these nearby voids is presented in two
variants. The larger sample which includes all galaxies falling within the
void boundaries, consists of 1354 objects. The subsample of inner void
galaxies, intended to study possible evolutionary peculiarities, includes
galaxies situated in the void interiors, with the distances to the nearest
luminous galaxy $D_{\rm NN} \geq$ 2.0~Mpc. It consists of 1088 objects.
\item
We present the preliminary statistical results for the Nearby Void Galaxy
sample. The $M_{\rm B}$ distribution of 1004 'inner' void sample with known
$B$-magnitudes has the median of \mbox{$\sim-15.1$} and extends from
\mbox{$M_{\rm B} = -19.5$} down to  \mbox{$M_{\rm B} \sim -7.1$}.
The absolute majority of void dwarfs are irregular galaxies and late-type
spirals. However, $\sim7$ per cent of the void sample are of early types:
dE/E-S0 galaxies with the wide range of $M_{\rm B}$. Most of them are well
isolated, and thus can represent an unusual sub-type of field/void early-type
galaxies.
\item
The new large sample of galaxies residing in the nearest voids opens the good
perspective of systematical studies of galaxy formation and evolution
as well as of the properties of voids themselves. These include the following
directions:
1) search for and study of the lowest mass void dwarfs, including candidates
to the so-called Very Young Galaxies; 2) study the origin and evolution of
early-type galaxies in voids; 3) search for appearance of cold accretion
as a driving mechanism of galaxy evolution; 4) study of void small-scale
structure as a probe of the possible role of Warm Dark Matter in its
formation,  5) study the
void galaxy motions from their void centers, as a mean to determine void
global properties, and several others.
\end{enumerate}

\section*{Acknowledgements}

The authors are pleased to acknowledge the support of this work through
the Russian Science Foundation grant No.~14-12-00965. We thank
I.D.~Karachentsev who read the manuscript and made valuable suggestions.
The authors are grateful to the anonimous referee for useful questions
and suggestions which helped to improve the paper content.
The usage of the HyperLEDA database \footnote{http://leda.univ-lyon1.fr}
as the
main source of the initial sample selection and an instrument for data checks
is greatly acknowledged.
This research has made use of the NASA/IPAC Extragalactic Database (NED)
which is operated by the Jet Propulsion Laboratory, California Institute
of Technology, under contract with the National Aeronautics and Space
Administration.
We also acknowledge the use of both the spectral and photomenric data
from the SDSS.
Funding for the Sloan Digital Sky Survey IV has been provided by
the Alfred P. Sloan Foundation, the U.S. Department of Energy
Office of Science, and the Participating Institutions. SDSS acknowledges
support and resources from the Center for High-Performance Computing
at the University of Utah. The SDSS web site is www.sdss.org.
SDSS is managed by the Astrophysical Research Consortium for the
Participating Institutions of the SDSS Collaboration.

\appendix

\section[]{On-line data for electronic version}
\label{App1}

\subsection{Tables of the Nearby Void Galaxy  catalog, 'inner' and 'outer'
parts}
\label{appen:gal.tables}

Here we present the compilation of main parameters for the whole
Nearby Void Galaxy sample divided into Table A1 and A2 with inner and
outer subsamples, respectively.  They both have the structure as
described in Subsection~\ref{ssec:void_gals}. Table A1 containes 1088 inner
void galaxies with Notes in the bottom of the table for part of individual
galaxies, mainly devoted to galaxy clustering with other void galaxies.
Table A2 containes 266 outer void galaxies with similar Notes in the bottom.

\subsection{3D views of the Nearby Voids from Table 1.}
\label{appen:voids3D}

In Figures A1--A6 of the Appendix we present 3D views in supergalactic
coordinates
$SGX, SGY, SGZ$ of all 25 voids from Table~2 in cubes with the side length of
50 Mpc, which includes the adopted main sphere with $R = 25$ Mpc. These views
allow one to get the
first impression of the void sizes and positions relative to the Local Sheet,
concentrated to the $SGZ = 0$ plane.

\subsection{Finding charts of the Nearby Void Galaxy catalog, the
'inner' and 'outer' parts}
\label{appen:f.charts}

We present the finding charts of the Nearby Void Galaxy sample in four
mosaics.
Figures A7--A65 are for finding charts of galaxies of the 'inner' part
(1088 objects with
$D_{\rm NN} \geq $2.0 Mpc) divided into those  falling within the SDSS
footprint sky regions (487 objects,  Figures A7--A25) and the remaining
601 'inner' galaxies residing outside the SDSS zone (Figures A26--A65),
for which we use digitized DSS1/DSS2 images.

Two more categories relate to the 'outer' void galaxies (266 objects with
$D_{\rm NN} < $2.0 Mpc). Figures A66--A70 show 110 galaxies falling
in the SDSS footprint, while Figures A71--A81  show the finding charts
of the remaining 156 'outer' objects similar to the 'inner' sample.

\label{lastpage}


\begin{thebibliography}{99}

\bibitem[\protect\citeauthoryear{Abazajian et al.}{2009}]{DR7}
	Abazajian K.N., Adelman-McCarthy J.K., Ag\"ueros M.A. et al.,
	 2009, ApJS, 182, 543

\bibitem[\protect\citeauthoryear{Angulo, Hahn, Abel}{2013}]{Angulo2013}
	Angulo R.E., Hahn O., \& Abel T., 2013, MNRAS, 434, 3337

\bibitem[\protect\citeauthoryear{Aragon-Calvo \& Szalay}{2013}]{AragonCalvoSzalay2013}
	Aragon-Calvo M.A. \& Szalay A.S., 2013, MNRAS, 428, 3409

\bibitem[\protect\citeauthoryear{Ashley, Marcum, \& Fanelli}{2017}]{Ashley17}
       Ashley T., Marcum P.M., \& Fanelli M.L., 2017, AJ, 153, 158

\bibitem[\protect\citeauthoryear{Beigu et al.}{2013}]{Beygu13}
     Beygu B., Kreckel K., van de Weygaert R., van der Hulst J.M.,
     van Gorkom J.H., 2013, AJ, 145, 120

\bibitem[\protect\citeauthoryear{Beigu et al.}{2017}]{Beygu17}
     Beygu B., Peletier R.F., van der Hulst J.M., Jarrett T.H.,
     Kreckel K., van de Weygaert R., van Gorkom J.H., Aragon-Calvo M.A.,
     2017, MNRAS, 464, 666

\bibitem[\protect\citeauthoryear{Bradford et al.}{2015}]{Bradford15}
    Bradford J.D., Geha M.C., Blanton M.R., 2015, ApJ, 809, 146


\bibitem[\protect\citeauthoryear{Brown et al.}{2015}]{Brown15}
  Brown W.R., Anderson J., Gnedin O.Y., Bond H.E., Geller M.J.,
  Kenyon S.J., 2015, ApJ, 804, 49

\bibitem[\protect\citeauthoryear{Chengalur, Pustilnik}{2013}]{CP13}
    Chengalur J.N., Pustilnik S.A., 2013, \mnras, 428, 1579


\bibitem[\protect\citeauthoryear{Chengalur, Pustilnik, Egorova}{2017}]
   {UGC3672A}
    Chengalur J.N., Pustilnik S.A., Egorova E.S., 2017, \mnras, 465, 2342

\bibitem[\protect\citeauthoryear{Colberg et al.}{2005}]{Colberg05}
 Colberg J.M., Sheth R.K., Diaferio A., Gao L., Yoshida N., 2005, \mnras, 360, 216

\bibitem[\protect\citeauthoryear{Colberg et al.}{2008}]{Colberg2008}
    Colberg J.M., Pearce F., Foster C., et al. 2008, \mnras, 387, 993

\bibitem[\protect\citeauthoryear{Colless et al.}{2003}]{2dFGRS}
   \mbox{Colless~M.}, \mbox{Peterson~B.}, \mbox{Jackson~C.}, et al. 2003,
  arXiv:0306581  (http://www.mso.anu.edu.au/2dFGRS)

\bibitem[\protect\citeauthoryear{Dawson et al.}{2016}]{SDSS-IV}
  Dawson K.S., Kneib J.-P., Percival W.J., et al., 2016, AJ, 151, 44


\bibitem[\protect\citeauthoryear{Ekta et al.}{2008}]{Ekta08}
   Ekta, Chengalur J.N., Pustilnik S.A., 2008, MNRAS, 391, 881

\bibitem[\protect\citeauthoryear{Elyiv et al.}{2013}]{Elyiv2013}
	Elyiv A.A., Karachentsev I.D., Karachentseva V.E., Melnyk O.V.,
	Makarov D.I., 2013, Astropysical Bulletin, 68, 1  (arXiv:1302.2369)

\bibitem[\protect\citeauthoryear{Fairall}{1998}]{Fairall98}
     Fairall A., 1998, Large-Scale Structures in the Universe, Wiley-Praxis,
     196 pp.

\bibitem[\protect\citeauthoryear{Flewelling et al.}{2016}]{PanSTARRS1}
  Flewelling H.A., Magnier E.A., Chambers K.C. et al., 2016, arXiv:1612.05243v2
  (submitted to ApJ)

\bibitem[\protect\citeauthoryear{Fliri \& Trujillio}{2016}]{Stripe82}
   Fliri J., Trujillio I., 2016, MNRAS, 456, 1359

\bibitem[\protect\citeauthoryear{Goldberg \& Vogeley}{2004}]{Goldberg2004}
    Goldberg D.M., Vogeley M.S., 2004, ApJ, 605, 1

\bibitem[\protect\citeauthoryear{Gottl\"ober et al.}{2003}]{Gottlober03}
    Gottl{\"o}ber S., Lokas E.L., Klypin A., Hoffman Y., 2003, MNRAS, 344, 715

\bibitem[\protect\citeauthoryear{Grogin, Geller}{1999}]{GG99}
    Grogin N.A., Geller M.J., 1999, AJ, 118, 2561

\bibitem[\protect\citeauthoryear{Grossi et al.}{2009}]{Grossi09}
    Grossi M., di Serego Alighieri S., Giovanardi C., Gavazzi G.,
    et al. 2009, A\&A, 498, 407

\bibitem[\protect\citeauthoryear{Hahn et al.}{2006}]{Hahn06}
	Hahn O., Porciani C.,  Carollo C.M., Dekel A.,
	2006,   MNRAS, 375, 489

\bibitem[\protect\citeauthoryear{Hahn et al.}{2007}]{Hahn07}
	Hahn O.,  Carollo C.M., Porciani C., Dekel A.,
	2007,   MNRAS, 381, 41

\bibitem[\protect\citeauthoryear{Hahn et al.}{2009}]{Hahn09}
	Hahn O., Porciani C., Dekel A., Carollo C.M., 2009,
	MNRAS, accepted (arXiv:0803.4211v2)


\bibitem[\protect\citeauthoryear{Haynes et al.}{2018}]{ALFALFA18}
     Haynes M.P., Giovanelli R., Kent B., et al., 2018,  arXiv:1805.11499

\bibitem[\protect\citeauthoryear{Hidding et al.}{2016}]{Hidding16}
  Hidding J., van de Weygaert R., Shandarin S.,
  The Zeldovich Universe: Genesis and Growth of the Cosmic Web, 2016,
  Proceedings of IAU,  IAU Symposium, V.308, pp.69--76

\bibitem[\protect\citeauthoryear{Hill et al.}{2010}]{Hill2010}
     Hill D.T., Driver S.P., Cameron E., Cross N., Liske J., Robotham A.,
     2010, MNRAS, 404, 1215

\bibitem[\protect\citeauthoryear{Hirschauer et al.}{2016}]{Hirschauer16}
     Hirschauer A.S., Salzer J.J., Skillman E.D., et al. 2016, ApJ, 822, 108

\bibitem[\protect\citeauthoryear{Hoeft et al.}{2006}]{Hoeft06}
     Hoeft M., Yepes G., Gottl{\"o}ber S., Springel V., 2006, MNRAS, 371, 401

\bibitem[\protect\citeauthoryear{Hoyle et al.}{2005}]{Hoyle05}
     Hoyle F., Rojas R.R., Vogeley M.S., Brinkmann J., 2005, ApJ, 600, 6018

\bibitem[\protect\citeauthoryear{Hoyle et al.}{2015}]{Hoyle15}
     Hoyle F., Vogeley M.S., Pan D., 2015, MNRAS, 426, 3041

\bibitem[\protect\citeauthoryear{Icke, van de Weygaert}{1987}]{IckeVanDeWeygaert1987}
    Icke V., van de Weygaert R., 1987, A\&A, 184, 16

\bibitem[\protect\citeauthoryear{Icke, van de Weygaert}{1991}]{IckeVanDeWeygaert1991}
    Icke V., van de Weygaert R., 1991, QJRAS, 32, 85

\bibitem[\protect\citeauthoryear{Izotov et al.}{2012}]{izotov2012}
    Izotov Y.I, Thuan T.X., Guseva N.G., 2012, A\&A, 546, 122

\bibitem[\protect\citeauthoryear{Izotov et al.}{2018}]{Izotov18}
    Izotov Y.I, Thuan T.X., Guseva N.G., Liss S.E., 2018, MNRAS, 473, 1956

\bibitem[\protect\citeauthoryear{Karachentsev}{2005}]{Karach2005}
    Karachentsev I.D., 2005, AJ, 129, 178

\bibitem[\protect\citeauthoryear{Karachentsev et al.}{2016}]{Kar2016}
   Karachentsev I.D., Chengalur J.N., Tully R.B., Makarova L.N.,
  Sharina M.E., Begum A., Rizzi L., 2016, AN, 337, 306

\bibitem[\protect\citeauthoryear{Karachentsev et al.}{2018}]{Virgo-front}
   Karachentsev I.D., Makarova L.N., Tully R.B., Rizzi L., Shaya E.,
   2018, ApJ, 858, 62

\bibitem[\protect\citeauthoryear{Karachentsev et al.}{1996}]{LGApex1996}
    Karachentsev I.D., Makarov D.I., 1996, AJ, 111, 794

\bibitem[\protect\citeauthoryear{Karachentsev et al.}{2008}]{LSCPairs}
    Karachentsev I.D., Makarov D.I., 2008, Astrophys. Bull., 63, 299

\bibitem[\protect\citeauthoryear{Karachentsev et al.}{2013}]{UNGC13}
	Karachentsev I.D., Makarov D.I., Kaisina E.I., 2013, AJ, 145, 101

\bibitem[\protect\citeauthoryear{Karachentsev, Nasonova \& Courtois}{2013}]{UMa2013}
Karachentsev I.D., Nasonova O.G., \& Courtois H., 2013, MNRAS, 429, 2264

\bibitem[\protect\citeauthoryear{Karachentsev et al.}{2011}]{LOG2011}
	Karachentsev I.D., Makarov D.I., Karachentseva V.E., Melnyk O.V.,
	2011, Astrophys. Bull., 65, 1

\bibitem[\protect\citeauthoryear{Karachentsev \& Nasonova}{2010}]{Virgo10}
	Karachentsev I.D., Nasonova O.G., 2010, MNRAS, 405, 1075
 The observed infall of galaxies towards the Virgo cluster


\bibitem[\protect\citeauthoryear{Karachentsev et al.}{2015a}]{LeoSpur15}
   Karachentsev I.D., Tully R.B., Makarova L.N., Makarov D.I., Rizzi L.,
   2015a, ApJ, 805, 144

\bibitem[\protect\citeauthoryear{Karachentsev et al.}{2015b}]{KKs3}
   Karachentsev I.D., Kniazev A.Y., Sharina M.E.,
   2015b, AN, 336, 707


\bibitem[\protect\citeauthoryear{Kniazev, Egorova, Pustilnik}{2018}]
  {Kniazev18}
 Kniazev A.Y., Egorova E.S., Pustilnik S.A., 2018, MNRAS, 479, 3842


\bibitem[\protect\citeauthoryear{Kreckel et al.}{2011}]{Kreckel11}
   Kreckel K., Platen E., Aragon-Calvo M.A., et al.
   2011, AJ, 141, 4

\bibitem[\protect\citeauthoryear{Kreckel et al.}{2012}]{Kreckel12}
   Kreckel K., Platen E., Aragon-Calvo M.A., van Gorkom J.H.,
   van de Weygaert R., van der Hulst J.M., Beygu B., 2012, AJ, 144, 16



\bibitem[\protect\citeauthoryear{Lindner et al.}{1996}]{Lindner96}
   Lindner U., Einasto M., Einasto J., et al. 1996, A\&A, 314, 1


\bibitem[\protect\citeauthoryear{Lupton et al.}{2005}]{Lupton05}
   \mbox{Lupton~R.,~et~al.~2005},
  http://www.sdss.org/dr5/algorithms\\/sdssUBVRITransform.html\#Lupton2005


\bibitem[\protect\citeauthoryear{Makarov \& Karachentsev}{2009}]{LSCTriplets}
   Makarov D.I., Karachentsev I.D., 2009, Astrophys. Bull., 64, 24

\bibitem[\protect\citeauthoryear{Makarov \& Karachentsev}{2011}]{LSCGroups}
  Makarov D.I., Karachentsev I.D., 2011, MNRAS, 412, 2498

\bibitem[\protect\citeauthoryear{Makarov et al.}{2014}]{HyperLEDA}
  Makarov D.I., Prugniel P., Terekhova N., Courtois H., Vauglin I.,
  2014, A\&A, 570, A13

\bibitem[\protect\citeauthoryear{Marchetti et al.}{2018}]{Marchetti18}
  Marchetti T., Contigiani O., Rossi E.M., Albert J.G., Brown A.G.A.,
  Sesana A.,  2018, MNRAS, 476, 4697

\bibitem[\protect\citeauthoryear{Matsuda, Shima}{1984}]{MatsudaShima1984}
  Matsuda T., Shima E., 1984, Prog. Theor. Phys., 71, 855

\bibitem[\protect\citeauthoryear{McIntyre}{2015}]{McIntyre2015}
McIntyre T.P., Henning P.A., Minchin R.F., Momjian E., Butcher Z.,
  2015, AJ, 150, 28

\bibitem[\protect\citeauthoryear{Moorman et al.}{2015}]{Moorman15}
  Moorman C.M., Prada F., Vogeley M.S., Hoyle F., Pan D.C.,
  Haynes M., Giovanelli R., 2015, ApJ, 810, 108

\bibitem[\protect\citeauthoryear{Moorman et al.}{2016}]{Moorman16}
  Moorman C.M., Moreno J., White A., Vogeley M.S., Hoyle F.,
  Haynes M., Giovanelli R., 2016, ApJ, 831, 118
  (arXiv:1601.04092)
 On the Star Formation Properties of Void Galaxies   2016ApJ...831..118M

\bibitem[\protect\citeauthoryear{Montero-Dorta, Prada}{2009}]{Montero09}
  Montero-Dorta A.D., Prada F., 2009, MNRAS, 339, 1106
 The SDSS DR6 luminosity functions of galaxies

\bibitem[\protect\citeauthoryear{Nasonova \& Karachentsev}{2011}]{LocalVoid11}
  Nasonova O.G.,  \& Karachentsev I.D., 2011, Astrophysics, 54, 1

\bibitem[\protect\citeauthoryear{Nasonova, de Freitas Pacheco, \& Karachentsev}{2010}]{Fornax11}
  Nasonova O.G., de Freitas Pacheco J.A., \& Karachentsev I.D., 2011, A\&A,
532, A104

\bibitem[\protect\citeauthoryear{Pan et al}{2012}]{Pan2012}
Pan D.C., Vogeley M.S.,  Hoyle F., Choi Y.-Y., Park C., 2012, MNRAS, 421, 926


\bibitem[\protect\citeauthoryear{Patiri et al.}{2006}]{Patiri06}
      Patiri S.G., Prada F., Holtzman J., Klypin A., Betancort-Rijo J.,
  2006, MNRAS, 372, 1710


\bibitem[\protect\citeauthoryear{Perepelitsyna et al.}{2014}]{PaperIV}
  Perepelitsyna Y.A.,  Pustilnik S.A., Kniazev A.Y.,
   2014, Astrophys. Bull., 69, 247  (arXiv:1408.0613) 

\bibitem[\protect\citeauthoryear{Prada et al.}{2003}]{Prada03}
   Prada F., Vitvitska M., Klypin A., et al.,  2003, ApJ, 598, 260

\bibitem[\protect\citeauthoryear{Pustilnik \& Tepliakova}{2011}]{PaperI}
 Pustilnik S.A., Tepliakova A.L., 2011, MNRAS, 415, 1188

\bibitem[\protect\citeauthoryear{Pustilnik \& Martin}{2016}]{Paper6}
 Pustilnik S.A., Martin J.-M., 2016, A\&A, 596, A86


\bibitem[\protect\citeauthoryear{Pustilnik, Kniazev \& Pramskij}{Pustilnik
	et al.}{2005}]{DDO68}
   Pustilnik S.A., Kniazev A.Y., Pramskij A.G., 2005, \aap, 443, 91

\bibitem[\protect\citeauthoryear{Pustilnik et al.}{2008}]{AndIV}
  Pustilnik S.A., Tepliakova A.L., Kniazev A.Y., Burenkov A.N.,
  2008, Astrophys. Bull., 63, 102 (arXiv:0712.4205)

\bibitem[\protect\citeauthoryear{Pustilnik et al.}{2010}]{J0926}
  Pustilnik S.A., Tepliakova A.L., Kniazev A.Y., Martin J.-M., Burenkov A.N.,
  2010, MNRAS, 401, 333

\bibitem[\protect\citeauthoryear{Pustilnik, Tepliakova \& Kniazev}{2011a}]
  {PaperII}
  Pustilnik S.A., Tepliakova A.L., Kniazev A.Y., 2011,
  Astrophys.~Bull., 66, 255 
(arXiv:1108.4850)

\bibitem[\protect\citeauthoryear{Pustilnik et al.}{2011b}]{void_LSBD}
   Pustilnik S.A., Martin J.-M., Tepliakova A.L., Kniazev A.Y.,
   2011, MNRAS, 417, 1335  

\bibitem[\protect\citeauthoryear{Pustilnik et al.}{2013}]{J0015}
  Pustilnik S.A., Martin J.-M., Lyamina Y.A., Kniazev A.Y.,
    2013, MNRAS, 432, 2224

\bibitem[\protect\citeauthoryear{Pustilnik et al.}{2016}]{Paper7}
  Pustilnik S.A.,  Perepelitsyna Y.A., Kniazev A.Y.,
    2016, MNRAS, 463, 670  

\bibitem[\protect\citeauthoryear{Rieder et al.}{2013}]{Rieder13}
    Rieder S., van de Weygaert R., Cautun M., Beygu B., Portegies Zwart S.,
    2013, MNRAS, 435, 222

\bibitem[\protect\citeauthoryear{Rieder et al.}{2016}]{Rieder16}
    Rieder S., van de Weygaert R., Cautun M., Beygu B., Portegies Zwart S.,
    2016, The Zeldovich Universe: Genesis and Growth of the Cosmic Web,
    Proc. of the IAU, IAU Symposium, Volume 308, pp. 575-579

\bibitem[\protect\citeauthoryear{Rizzi et al.}{2017}]{Rizzi2017}
    Rizzi L., Tully R.B., Shaya E.J., Kourkchi E., Karachentsev I.D.,
   2017, ApJ, 835, 78

\bibitem[\protect\citeauthoryear{Rojas et al.}{2004}]{Rojas04}
  Rojas R.R., Vogeley M.S., Hoyle F., Brinkmann J., 2004, ApJ, 617, 50

\bibitem[\protect\citeauthoryear{Rojas et al.}{2005}]{Rojas05}
 Rojas R.R., Vogeley M.S., Hoyle F., Brinkmann J., 2005, ApJ, 624, 571

\bibitem[\protect\citeauthoryear{Salzer}{1992}]{Salzer92}
    Salzer J.J.,   1992, AJ, 103, 385

\bibitem[\protect\citeauthoryear{Salzer et al.}{1991}]{Salzer91}
    Salzer J.J.,  di Serego Alighieri S., Matteucci F., Giovanelli R.,
  Haynes M.P., 1991, AJ, 101, 1258


\bibitem[\protect\citeauthoryear{Schlafly \& Finkbeiner}{2011}]{SF2011}
    Schlafly E.F.,  Finkbeiner D.P.,  2011, ApJ, 737, a.i. 103

\bibitem[\protect\citeauthoryear{Sheth \& van de Weygaert}{2004}]{Sheth2004}
   Sheth R.K., van de Weygaert R., 2004, MNRAS, 350, 517

\bibitem[\protect\citeauthoryear{Skrutskie et al.}{2006}]{2MASS}
   Skrutskie M.F., Cutri R.M., Stiening R., et al., 2006, AJ,  131, 1163

\bibitem[\protect\citeauthoryear{Sorgho et al.}{2018}]{MHONGOOSE}
    Sorgho A., Carignan C., Pisano D.J., et al., 2018MNRAS.tmp.2656S

\bibitem[\protect\citeauthoryear{Sorrentino, Antonuccio-Delogu \& Rifatto}
 {2006}]{Sorrentino06}
 Sorrentino G., Antonuccio-Delogu V., Rifatto A., 2006, A\&A, 460, 673

\bibitem[\protect\citeauthoryear{Staveley-Smith et al.}{2016}]{Staveley2016}
   Staveley-Smith L., Kraan-Korteweg R.C., Schr\''oder A.C., Koribalski B.S.,
   Stewart I.M., Heald G., 2016, AJ, 151, article id.~52

\bibitem[\protect\citeauthoryear{Stierwalt et al.}{2009}]{Stierwalt09}
 Stierwalt S., Haynes M.P.,  Giovanelli R., Kent B.R.,  Martin A.M.,
  Saintonge A., Karachentsev I.D., Karachentseva V.E.,
   2009, AJ, 138, 338

\bibitem[\protect\citeauthoryear{Tikhonov, Karachentsev}{2006}]{TK06}
   Tikhonov A.V., Karachentsev I.D., 2006, ApJ, 653, 969

\bibitem[\protect\citeauthoryear{Tikhonov et al.}{2009}]{Tikhonov09}
   Tikhonov A.V., Gottlober S., Yepes G., Hoffman Y.;
   2009, MNRAS, 399, 1611

\bibitem[\protect\citeauthoryear{Tully et al.}{2008}]{Tully2008}
    Tully R.B., Shaya E.J., Karachentsev I.D., Courtois H.M., Kocevski D.D.,
    Rizzi L., Peel A.,   2008, \apj, 676, 184

\bibitem[\protect\citeauthoryear{Tully et al.}{2009}]{Tully09}
  Tully R.B., Rizzi L., Shaya E.J., Courtois H.M., Makarov D.I., Jacobs B.A.,
  2009, AJ, 138, 323

\bibitem[\protect\citeauthoryear{Tweed et al.}{2018}]{Tweed2018}
  Tweed D.P., Mamon G.A., Thuan T.X., Cattaneo A., Dekel A., Menci N., Calura F., Silk J.,
  2018, MNRAS, accepted (arXiv:1802.09530)

\bibitem[\protect\citeauthoryear{van de Weygaert}{2016}]{Weygaert16}
  van de Weygaert R.,
  The Zeldovich Universe: Genesis and Growth of the Cosmic Web, Proceedings of IAU,
  2016, IAU Symposium, V.308, pp.493--523

\bibitem[\protect\citeauthoryear{van de Weygaert, Icke}{1989}]
{VanDeWeygaertIcke1989}
    van de Weygaert R., Icke V., 1989, A\&A, 213, 1

\bibitem[\protect\citeauthoryear{van der Weygaert et al.}{2009}]{vandeWey09}
 van de Weygaert R., Platen E., Tigrak E., Hidding J., van der Hulst J.M.,
 Arag\'on-Calvo M.A., Stanonik K., van Gorkom J.H., 2009, in ASP Conf.Ser.,
 421, 99 


\bibitem[\protect\citeauthoryear{Yang et al.}{2015}]{Yang2015}
Yang L.F., Neyrinck M.C., Aragon-Calvo M.A., Falck B., Silk J., 2015, MNRAS, 451, 3606


\bibitem[\protect\citeauthoryear{Zibetti, Charlot, Rix}{2009}]{Zibetti09}
Zibetti S., Charlot S., and Rix H.-W., 2009, MNRAS, 400, 1181

\end{thebibliography}
\end{document}